\documentclass[12pt,twoside]{article}

\def\TheAuthor{Joachim Wuttke}
\def\TheTitle{Macromolecular Dynamics}
\def\TheHeading{Macromolecular Dynamics}
\def\TheInstitute{J\"ulich Centre for Neutron Science at FRM II}
\def\TheAddress{Forschungszentrum J\"ulich GmbH}
\def\ThePart{IFF Spring School 2011, B}
\def\TheChapter{4}

\usepackage{ferienschule_11}
\usepackage{upgreek}

\newcommand{\D}[1]{\textit{#1}}
\renewcommand{\v}[1]{\mbox{\boldmath$#1$}}
\newcommand{\V}[1]{\underline{#1}}
\newcommand{\M}[1]{\underline{\underline{#1}}}
\newcommand{\etal}[1]{\textit{et al.}}

\newcommand{\DS}{\displaystyle}

\begin{document}
\parindent=1.6em

\thispagestyle{empty}

\normalsize\rmfamily\strut\\[8ex]

\noindent
\textbf{\Huge Macromolecular Dynamics}\\[5ex]
{\Large An introductory lecture\\[4ex]
Joachim Wuttke\\[3ex]
J\"ulich Centre for Neutron Science at FRM II\\[.8ex]
Forschungszentrum J\"ulich GmbH}\vfill

\noindent
This text appeared in:\\
Macromolecular Systems in Soft and Living Matter,
Lecture notes of the 42nd IFF Spring School 2011,
edited by
Jan K.~G.~Donth, Gerhard Gompper, Peter R.~Lang,
Dieter Richter, Marisol Ripoll, Dieter Willbold, Reiner Zorn.\\
Schriften des Forschungszentrums J\"ulich, ISBN 978-3-89336-688-0,
J\"ulich 2011.

\newpage

\setcounter{page}{1}
\tableofcontents     

\newpage

\section{Systems and States}

This lecture on polymer dynamics\index{polymer dynamics}
 primarily addresses the molecular motion in melts and solutions.
Other states like rubber or glass will be refered to briefly.
Therefore,
let us start by sorting out the following states of
macromolecular matter:


A \D{polymer melt}\index{polymer melt}\index{polymer melt}
 is basically a liquid.
However, in contrast to a \D{simple} liquid,
its constituent molecules are flexible,
and therefore their center-of-mass motion
is usually accompanied by conformational changes.
The dynamics is further complicated by the entanglement\index{entanglement}
 of neighbouring chain molecules.
As a result, polymer melts are highly viscous and non-Newtonian liquids.
The prime example for a naturally occuring polymer melt is
latex\index{latex},
a milky fluid rich in cis-polyisoprene\index{polyisoprene}
that is found in many plants,
particularly abundant in the rubber tree. 

A \D{rubber}\index{rubber}
 (or \D{elastomer}\index{elastomer}) is obtained from a polymer melt
by a chemical process called \D{vulcanization}\index{vulcanization}
that creates permanent inter-chain links.
These cross-links prevent long-ranged diffusion and viscous flow.
On short time and length scales, however,
the polymer segments can move as freely as in a melt,
which explains the characteristic elasticity of the rubber state.
Everyday examples: erasers, car tyres.
An unvulcanised polymer melt is said to be in a \D{rubber state}
if the entanglement is so strong that the chains do not flow
on the time scale of observation.

\D{Plastic}\index{plastic}
 materials are either thermosetting or thermoplastic.

A \D{thermosetting plastic}\index{thermosetting}
is obtained by irreversibly creating strong chemical inter-chain links.
Examples: bakelite\index{bakelite}, duroplast\index{duroplast},
epoxy resin\index{epoxy resin}\index{resin}.

\D{Thermoplastic}\index{thermoplastic} materials are
obtained in a reversible manner by cooling a polymer melt.
They can be either amorphous or semi-crystalline.

An \D{amorphous} solid\index{amorphous solid}
is also called a \D{glass}\index{glass},
provided that it can be reversibly transformed into a liquid.
This transformation is called the 
\D{glass transition}\index{glass transition};
it occurs at a loosely defined
temperature~$T_{\rm g}$\index{T@$T_{\rm g}$}.
Examples:
poly(methyl methacrylate)\index{polymethylmetacrylate@poly(methyl
   methacrylate)} (PMMA, plexiglas),
\index{plexiglas|see{poly(methyl methacrylate)}}
polystyrene\index{polystyrene},
polycarbonate\index{polycarbonate}.

In \D{semicrystalline}\index{semicrystalline} polymers,
chains run through ordered and disordered domains.
The ordered, crystalline domains exist
up to a melting point $T_{\rm m}$.\index{T@$T_{\rm m}$}
They provide strong links between different chains.
In the temperature range between $T_{\rm g}$ and $T_{\rm m}$,
the disordered domains provide flexibility.
Example: polyethylene\index{polyethylene}
 ($T_{\rm g}\simeq-100\ldots-70^\circ$C, 
$T_{\rm m}\simeq110\ldots140^\circ$C), used for foils and bags.
Most other thermoplastics are typically employed below $T_{\rm g}$.
Example: polyethylene terephthalate\index{polyethylene terephthalate}
 (PET, $T_{\rm g}\simeq75^\circ$C,
$T_{\rm m}\simeq260^\circ$C), used for bottles and textile fibers.

In a \D{polymer solution}\index{polymer solution}
the dilute macromolecules are always flexible.
Therefore they behave as in a pure polymer melt,
except that interactions between segments are modified
by the presence of a solvent.


\section{Brownian Motion}\label{SBrown}

Almost all of polymer dynamics can be described
by means of \D{classical mechanics}.
Quantum mechanics seems relevant only for the explanation of certain
low temperature anomalies that are beyond the scope of this lecture.
Since we aim at understanding macroscopically relevant \D{average} properties
of a large number of molecules,
the appropriate methods are provided by \D{statistical mechanics}.

Some fundamental methods of classical statistical mechanics have
been originally developped for the description of
\D{Brownian motion}\index{Brownian motion} \cite{Dup05}.
In this introductory section, we will review two of them:
the \D{Langevin equation} for 
the evaluation of particle trajectories,
and
the \D{Smoluchowski equation} for a higher-level modelling
in terms of probability densities.
Both methods are ultimately equivalent,
but depending on the application there can be a huge computational
advantage in using the one or the other.\footnote
{The following is standard material,
covered by many statistical mechanics textbooks
and by a number of monographs.
For Brownian motion, see e.~g.\ \cite{Maz02},
for the Langevin equation, \cite{CoKW04}.}

\subsection{The Langevin equation}\label{SLangevin}

We consider a colloidal (mesoscopic) particle $P$ suspended in a liquid $L$.
Molecules of $L$ frequently collide with $P$,
thereby exerting a random force $\v{F}(t)$.
On average the impacts from opposite directions cancel, so that
\begin{equation}\label{Eramom1}
  \langle \v{F}(t) \rangle = 0.
\end{equation}
For this reason it was long believed that $L-P$ collisions cannot
be responsible for the random motion of $P$.
By noting the analogy with the fluctuations of good or bad luck in gambling,
Smoluchowski (1906) showed that this argument is fallacious:
After $n$ collisions,
the expectation value of total momentum transfer is indeed~0,
but for each single history of $n$ collisions one
can nevertheless expect with high probability
that the momentum transfer has summed up to a value different from~0.

This \D{random walk}\index{random walk} argument was put on a firm base
by Langevin \index{Langevin equation} (1908).\footnote
{K.~Razi Naqvi [arXiv:physics/052141v1] contests this received opinion,
arguing that Langevin's ``analysis is at best incomplete,
and at worst a mere tautology.''}
Let us write down the Newtonian equation of motion for 
the center-of-mass coordinate $\v{r}(t)$ of $P$:
\begin{equation}\label{ELv1}
  m\partial^2_t \v{r} = - \zeta\partial_t\v{r} + \v{F},
\end{equation}
or for the velocity $\v{v}=\partial_t\v{r}$:
\begin{equation}\label{ELv2}
  m\partial_t \v{v} = - \zeta\v{v} + \v{F}.
\end{equation}
If $P$ is a spherical particle of radius $a$, and $L$ has
a shear viscosity\index{shear viscosity}~$\eta_{\rm s}$,
then the \D{friction coefficient}\index{friction coefficient}
is (Stokes 1851)\index{Stokes law} \cite[\S~20]{LL6}
\begin{equation}\label{EStokes}
  \zeta = 6\pi\eta_{\rm s} a.
\end{equation}
Eq.~(\ref{ELv2}) can be easily integrated:
\begin{equation}\label{Evt}
  \v{v}(t) = 
     m^{-1}\int_{-\infty}^t\!{\rm d}t'\,\v{F}(t'){\rm e}^{-(t-t')/{\tau_m}}.
\end{equation}
$P$ may be said to have a \D{memory}\index{memory} of past collisions
that decays with a \D{relaxation time}
\begin{equation}\label{Etaum}
  \tau_m:=\frac{m}{\zeta}.
\end{equation}
To compute averages like $\langle\v{v^2}\rangle$ 
we must now specify the expectation value of
random force correlation.
We assume
that different Cartesian components of $\v{F}$
are uncorrelated with each other,
$\langle F_\alpha F_\beta\rangle=0$ for $\alpha\ne\beta$,
and that random forces acting at different times $t,t'$
are uncorrelated.
The second moment of $\v{F}$ then takes the form
\begin{equation}\label{Eramom2}
  \langle F_\alpha(t)F_\beta(t')\rangle = A\delta_{\alpha\beta} \delta(t-t')
\end{equation}
where the delta function is a convenient approximation for a memory
function that extends over no more than the duration of one $L-P$ collision.

By virtue of (\ref{Eramom1}),
the average velocity is $\langle\v{v}\rangle=0$.
However, as argued above, for each single history of $\v{F}$ one
can expect that the integral in (\ref{Evt}) results in a nonzero velocity.
After a short calculation we find that the
second moment has the time-independent value
\begin{equation}\label{Ev2}
     \langle\v{v}^2(t)\rangle
  =  \frac{3A}{2m\zeta}.
\end{equation}
On the other hand, this moment 
is well known from the equipartition theorem
\begin{equation}\label{Eequip}
  \frac{m}{2}\langle\v{v^2}\rangle = \frac{3}{2}k_{\rm B}T,
\end{equation}
so that we obtain the coefficient
$ 
  A=2k_{\rm B}T\zeta
$ 
that was left unspecified in (\ref{Eramom2}).

In the next step,
we can compute the
\D{mean squared displacement}\index{mean squared displacement}
of $P$ within a time span~$t$,
which we will abbreviate as
\begin{equation}\label{Er2t}
  \langle r^2(t) \rangle :=
  \left\langle {\left[\v{r}(t)-\v{r}(0) \right]}^2 \right\rangle.
\end{equation}
We determine $\v{r}(t)$ by explicit integration of~(\ref{Evt}),
and after \D{some} calculation \cite{CoKW04} we find
\begin{equation}\label{Er2Langevin}
  \langle r^2(t) \rangle =
  6Dt-6D\tau_m\left(1-{\rm e}^{-t/\tau_m}\right)
\end{equation}
where we have introduced
\begin{equation}\label{EEinstein}
  D := \frac{k_{\rm B}T}{\zeta}.
\end{equation}
From the long-time limit $\langle r^2(t) \rangle\simeq 6Dt$,
we identify $D$ as the \D{diffusion coefficient}\index{diffusion coefficient}
(Sutherland 1905, Einstein 1905).
In the opposite limit $t\ll\tau_m$, 
ballistic motion is found:
$\langle r^2(t) \rangle\doteq\langle{(\v{v}(0)t)}^2\rangle$.

\subsection{The Smoluchowski equation}

Alternatively,
Brownian motion can be analysed in terms of the space-time distribution
$\rho(\v{r},t)$ of suspended particles.
The continuity equation is
\begin{equation}
  \partial_t \rho + \v{\nabla}\v{j}=0.
\end{equation}
As usual, the current $\v{j}$ has a diffusive component (Fick 1855)
\index{Fick's law}
\begin{equation}
  \v{j}^{\rm diff}=-D\v{\nabla}\rho.
\end{equation}
We now assume that a potential $U(\v{r})$ is acting upon the particles.
In the stationary state, the force $-\v{\nabla}U$ is just
cancelled by the friction $-\zeta\vec{v}$.
Accordingly, the drift component of $\v{j}$ is
\begin{equation}
  \v{j}^{\rm drift}=\rho\v{v} =-\zeta^{-1}\rho\v{\nabla} U.
\end{equation}
Collecting everything,
we obtain the \D{Smoluchowski equation}\index{Smoluchowski equation} (1915)
\begin{equation}
  \partial_t \rho = D\v{\nabla}^2\rho + \zeta^{-1}\v{\nabla}(\rho\v{\nabla} U).
\end{equation}
For a velocity distribution or for the more general case of
a phase-space distribution
it is known as the
\D{Fokker-Planck equation}\index{Fokker-Planck equation} (1914/1917).

In the stationary state $\partial_t\rho=0$,
we expect to find a Boltzmann distribution $\rho\propto(-U/k_{\rm B}T)$.
It is easily verified that this holds provided $D$ and $\zeta$ are
related by the \D{Sutherland-Einstein relation}%
\index{Einstein relation|see{Sutherland-Einstein relation}}%
\index{Sutherland-Einstein relation}~(\ref{EEinstein}).
Combined with~(\ref{EStokes}),
the \D{Stokes-Einstein relation}\index{Stokes-Einstein relation}
\begin{equation}\label{EEinsteinStokes}
  D = \frac{k_{\rm B}T}{6\pi\eta_{\rm s} a}
\end{equation}
is obtained.
Surprisingly,
this relation holds not only for mesoscopic suspended particles,
but even for the motion of one liquid molecule among others
(the \D{hydrodynamic radius}\index{hydrodynamic radius}~$a$ is
then treated as an adjustable parameter).
Only in the highly viscous supercooled state,
the proportionality $D\propto T/\eta_{\rm s}$ is found to break down.

\section{Segmental Relaxation and the Glass Transition}

\index{segmental dynamics}\index{glass transition}

\subsection{The glass transition}

It is generally believed that the ground state
of condensed matter is always an ordered one.
However, in many supercooled liquids\index{supercooled liquid}
this ordered state is kinetically inaccessible.
Crystallization through homogeneous nucleation\index{nucleation} requires
the spontaneous formation of nuclei of a certain critical size.
If the crystalline unit cell is complicated and the energy gain small,
then this critical size is rather larger.
If at the same time a high viscosity entails a low molecular mobility,
then the formation of a stable nucleus may be so improbable that
crystallization is practically excluded.
On further supercooling the liquid becomes
an amorphous solid,\index{amorphous solid}
 a \D{glass}.\index{glass}

Glass formation is observed in quite different classes of materials:
in covalent networks like quartz glass (SiO$_2$) or industrial glass
(SiO$_2$ with additives),
in molecular liquids,
in ionic mixtures,
in aqueous solutions, and others.
In polymers, glass formation is the rule, not the exception.
As said in Sect.~1,
some plastic materials are completely amorphous;
in others crystalline domains are intercalated between amorphous regions.

\begin{figure} 
\centering
\includegraphics[width=0.4\textwidth]{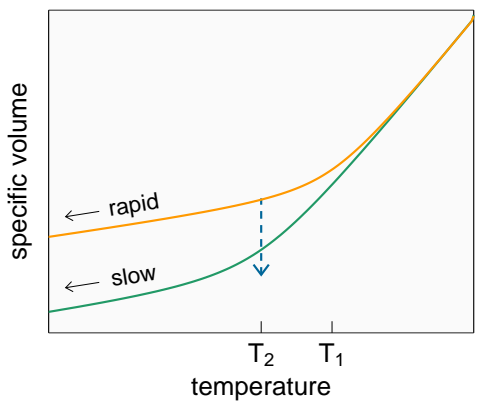}
\caption{Temperature dependence of the specific volume 
of a glass forming system.
Depending on the cooling rate,
the cross-over from liquid-like to glass-like slopes
is observed at different temperatures,
and it results in different glass states.
If the system is cooled rapidly to $T_2$ and then kept at that temperature,
it relaxes towards states that are normally obtained by slow cooling.
Similar behavior is found for the enthalpy.}
\label{Fglatra-V-T}
\end{figure} 

The \D{glass transition} is \D{not} a usual thermodynamic phase transition
of first or second order,
as can be seen from the temperature dependence of enthalpy or
density (Fig.~\ref{Fglatra-V-T}):
the cross-over from liquid-like to solid-like slopes is smooth,
and it depends on the cooling rate.
When the cooling is interrupted within the transition range,
\D{structural relaxation}\index{structural relaxation} 
towards a denser state is observed.
This shows that glass is not an equilibrium state:
on the contrary, it can be described as a liquid
that has fallen out of equilibrium.\footnote
{You may object that the supercooled liquid is already out of
equilibrium --- with respect to crystallization.
However, the crystalline ground state is a null set in phase space:
beyond nucleation it has no impact upon the dynamics of the liquid.
The glass transition, in contrast, can be described as
an ergodicity breaking in phase space.\index{ergodicity}\par
The thermodynamics of the glass transition is subject of ongoing debate.
A recent attempt to clarify confused concepts is \cite{LeNi08}.}

\subsection{Structural relaxation}\label{Sarx}

\begin{figure} 
\centering
\includegraphics[width=0.46\textwidth]{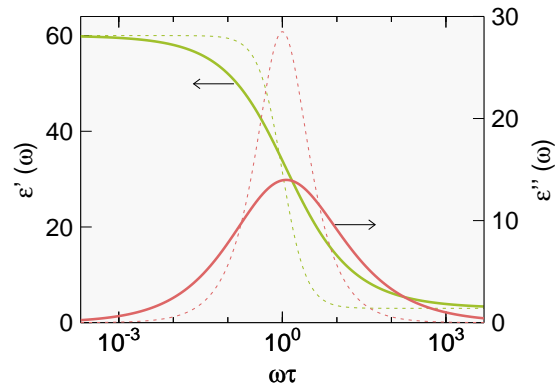}
\caption{Dielectric spectrum in the viscous liquid or rubber state.
As long as molecular dipoles are able to follow an external 
electric field modulation,
the permittivity $\epsilon(\omega)=\epsilon'(\omega)+i\epsilon''(\omega)$
has the elevated low-frequency value
of a polar liquid, here $\epsilon_0=60$.
On the other hand, when probed with a high-frequency field modulation,
dipoles appear frozen,
resulting in the low permittivity of a solid, here $\epsilon_\infty=3$.
As required by the Kramers-Kronig relation,
the dispersion of $\epsilon'(\omega)$
is accompanied by a dissipation peak in $\epsilon''(\omega)$.
The dispersion step and the loss peak are {\rm stretched} when compared to
 Debye's Eq.~(\ref{EsusDeb}) (dashed).}
\label{Fa-relax-havneg}
\end{figure} 

Relaxation can also be probed by applying a periodic perturbation.
This is done for instance in dielectric spectroscopy
(Fig.~\ref{Fa-relax-havneg}).
For reference,
the most elementary theory of dielectric relaxation
\index{dielectric spectroscopy}
(Debye 1913) is derived in Appendix~\ref{ADebye}.
\index{Debye relaxation}
It leads to a complex permittivity
\begin{equation}\label{EDebye}
  \epsilon(\omega) = \epsilon_\infty
     + \frac{(\epsilon_0-\epsilon_\infty)}
            {\left(1-(i\omega\tau)^\alpha\right)^\gamma}
\end{equation}
with $\alpha=\gamma=1$.
Empirically the cross-over from liquid-like to solid-like
behavior extends over a much wider frequency range
than this formula predicts: it is \D{stretched}\index{stretching}.
Often, this stretching is well described 
if the exponents anticipated in Eq.~(\ref{EDebye}) are allowed to
take values below~1
(Cole, Cole 1941,\index{Cole-Cole function}
 Cole, Davidson 1951,\index{Cole-Davidson function}
 Havriliak, Negami 1967).\index{Havriliak-Negami function}
Alternatively (Williams, Watts 1970),\index{Williams-Watts function}
$\epsilon(\omega)$ can be fitted by the Fourier transform of
the \D{stretched exponential} \index{stretched exponential function}
 function
 \begin{equation}\label{EKohl}
   \Phi_{\rm K}(t) = \exp(-(t/\tau)^\beta)
 \end{equation}
originally introduced to describe relaxation
in the time domain (R.~Kohlrausch 1854,
F.~Kohl\-rausch 1863).\index{Kohlrausch 
  function|see{stretched exponential function}}

Generally, the relaxation time $\tau$ depends much stronger
on temperature than the stretching exponents $\alpha,\gamma$ or $\beta$.
This can be expressed as a \D{scaling law}\index{scaling},
often called
\D{time-temperature superposition 
  principle}\index{time-temperature superposition}:
Permittivities $\epsilon(\omega;T)$
measured at different temperatures~$T$
can be rescaled with times $\tau(T)$ to fall onto a common 
\D{master curve}\index{master curve} $\hat\epsilon(\tau(T)\omega)$.

\begin{figure} 
\strut
\hfill
\includegraphics[width=0.44\textwidth]{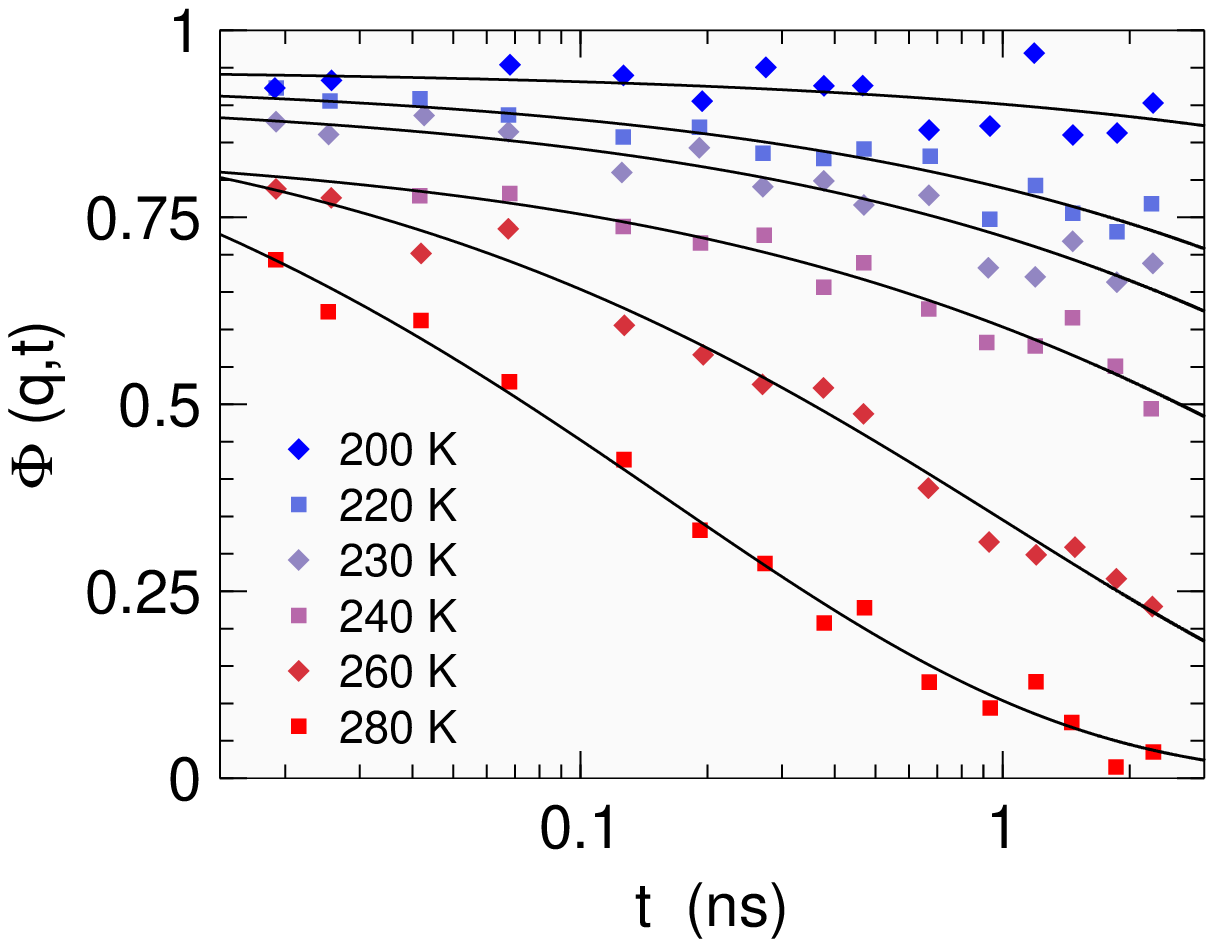}
\hfill
\includegraphics[width=0.44\textwidth]{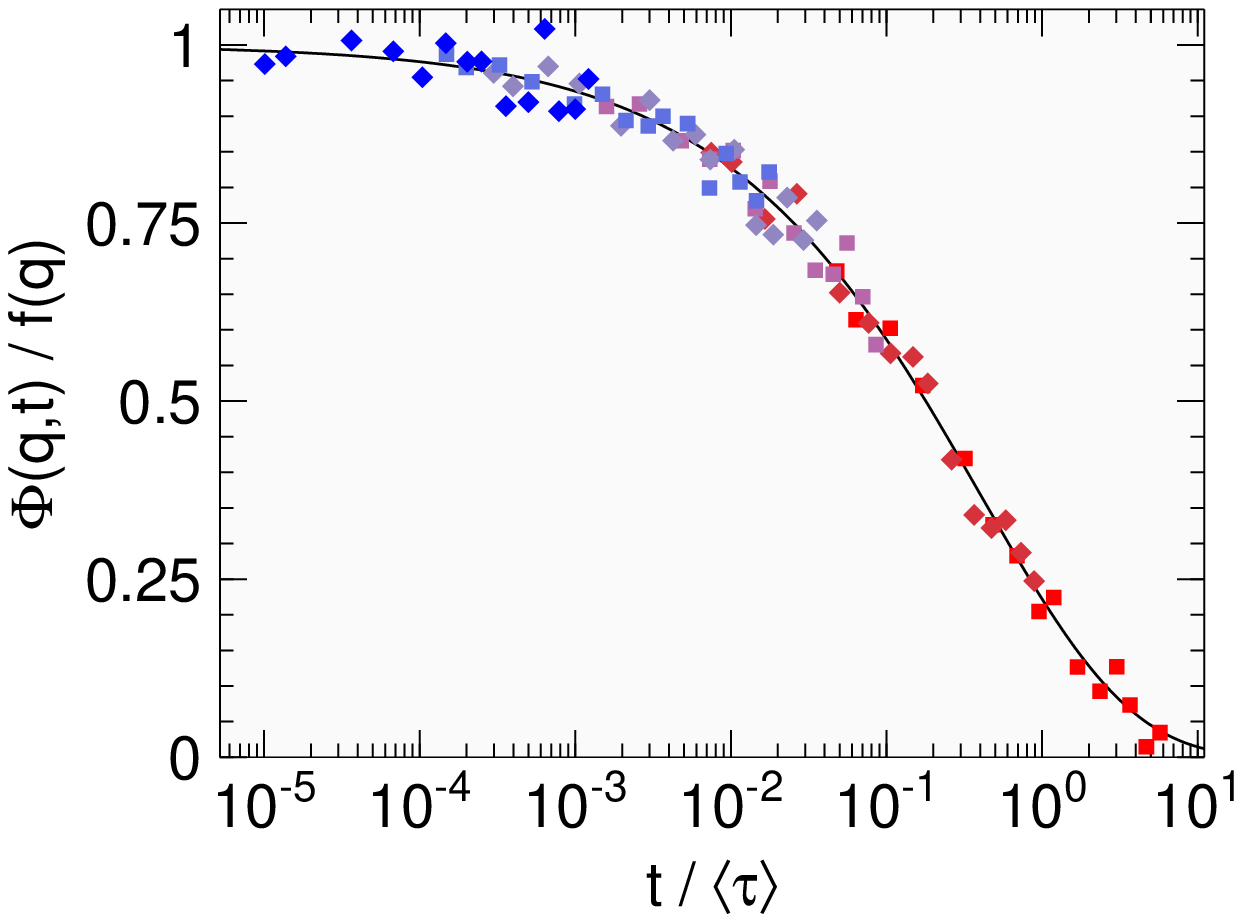}
\hfill
\strut
\caption{(a) Correlation function $\Phi(q,t)$ of deuterated
polybutadiene\index{polybutadiene},
measured by neutron spin-echo at $q=1.5$~\AA$^{-1}$ (close to
the maximum of $S(q)$).
Solid lines: fits with $f_q\exp(-(t/\tau)^\beta)$ with
fixed~$\beta=0.45$.
(b) Master curve.
Data from~\cite{RiFF88}.}
\label{FRiFF88}
\end{figure} 

Quite similar results are obtained
for other response functions.
For instance, the longitudinal mechanic 
\index{longitudinal modulus}
modulus,
a linear combination of the more fundamental shear
\index{shear modulus}
and bulk 
\index{bulk modulus}
moduli,
can be probed by ultrasonic propagation and attenuation
\index{ultrasound}
in a kHz\ldots MHz range,
or by light scattering 
\index{Brillouin-Mandelstam scattering}
\index{light scattering}
\index{hypersound}
in a ``hypersonic'' GHz range (Brillouin 1922, Mandelstam 1926).

Particularly interesting is the pure shear modulus~$G(\omega)$,
which can be probed by torsional spectroscopy.
Since a liquid cannot sustain static shear, there is no constant term
in the low-frequency expansion
\begin{equation}\label{EGliq}
   G(\omega) = i\eta\omega + {\cal O}(\omega^2).
\end{equation}
The coefficient $\eta$ is the
macroscopic \D{shear viscosity}\index{shear viscosity}.
If $G(\omega)$ obeys time-temperature superposition,
then $\eta(T)$ is proportional to a shear relaxation time $\tau_{\eta}(T)$.

By virtue of the fluctuation-dissipation theorem (Appendix~\ref{ALinResp}),
relaxations can also be studied \D{in} equilibrium,
via correlation functions that can be measured in
inelastic scattering experiments.
Fig.~\ref{FRiFF88} shows the normalized correlation function
\begin{equation}\label{EPhi}
\Phi(q,t):=S(q,t)/S(q,0)
\end{equation}
of a glass-forming polymer,
measured in the time domain by neutron spin echo.\index{neutron spin echo}
Scaling is demonstrated by construction of a master curve,
stretching by fits with a Kohlrausch function.

\subsection{Relaxation map and secondary relaxation}\label{Sbrx}

\begin{figure} 
\centering
\includegraphics[width=0.48\textwidth]{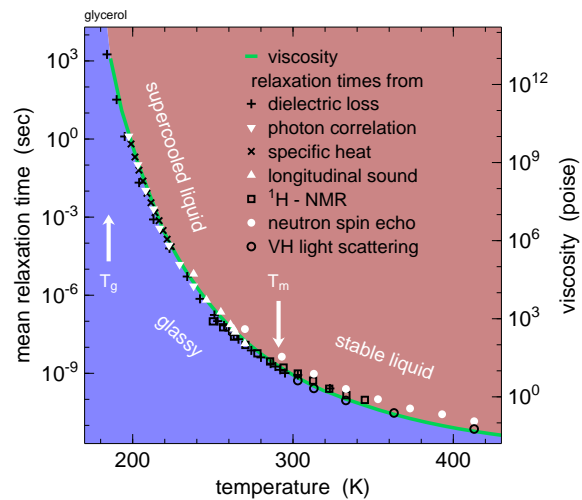}
\caption{Relaxation times in glycerol,
determined by various spectroscopies.
This plot can be read as a {\rm dynamic phase diagram:}
\index{dynamic phase diagram}
whether the material reacts like a liquid or a solid
depends not only on temperature, but also on the time scale of observation.
Adapted from~\cite{Wut00d}.}
\label{Farx-tau+}
\end{figure} 

At this point it is interesting to compare the outcome of different
experimental methods.
It is found that the susceptibility master curve is not universal;
different spectroscopies generally yield different stretching exponents.
Relaxation times may vary by factors~2 or more.
However,
with good accuracy all relaxation times have the same
\D{temperature dependence}.
This is demonstrated in Fig.~\ref{Farx-tau+} for a particularly well studied
molecular liquid.
Upon cooling from 290 to 190~K, the relaxation times increase in parallel
by almost 12 decades.
When a viscosity of $10^{13}$~Poise ($10^{12}$~Pa\,s) or
a relaxation time of about $10^3$~s is reached,
relaxation falls out of equilibrium on the time scale of a human observer.
As anticipated above, this is the glass transition\index{glass transition}.
All relaxation modes that follow the temperature dependence of viscosity
are conventionally called
$\upalpha$ relaxation.\index{$\upalpha$ relaxation}

The temperature dependence of relaxation times is
often discussed with reference to the simplest physical model
that comes to mind:
thermally activated jumps over an energy barrier~$E_{\rm A}$,
\begin{equation}\label{EArrh}
  \tau = \tau_0 \exp \frac{E_{\rm A}}{k_{\rm B}T}
\end{equation}
(van t'Hoff 1884, Arrhenius 1889).
Therefore,
data are often plotted as $\log\tau$ versus $1/T$ or $T_{\rm g}/T$
so that a straight line is obtained if (\ref{EArrh})~holds.
The $\upalpha$ relaxation in glass forming liquids, however,
almost never follows~(\ref{EArrh});
its trace in the Arrhenius plot\index{Arrhenius plot}
 is more or less concave.
Many fitting formul\ae\ have been proposed;
for most applications it is sufficient to extend (\ref{EArrh})
by just one more parameter,
as in the Vogel-Fulcher-Tammann equation
(for polymers also named after Williams, Landel, Ferry)
\begin{equation}\label{EVFT}
  \tau = \tau_0 \exp \frac{A}{T-T_0}.
\end{equation}
The singularity $T_0$ lies below~$T_{\rm g}$,
and it is unclear whether it has any physical meaning.

\begin{figure} 
\centering
\includegraphics[width=0.48\textwidth]{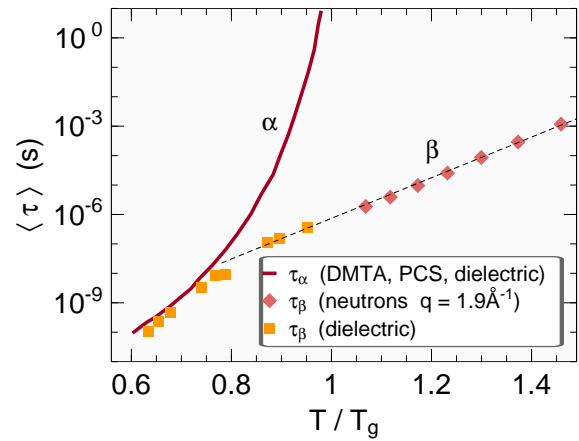}
\caption{$\upalpha$ and $\upbeta$ relaxation times 
of cross-linked polyurethane\index{polyurethane},
obtained by differential mechanical-thermal analysis (DMTA),
photon correlation spectroscopy (PCS),
dielectric spectroscopy and neutron scattering.
Data from~\cite{LeFP00}.}
\label{Fabrx}
\end{figure}

In many materials,
there is more than just one relaxation process.
If the additional, \D{secondary}\index{secondary relaxation}
 processes are faster than $\upalpha$~relaxation,
they are conventionally labelled $\upbeta$, $\upgamma$, \ldots.
\index{$\upbeta$ relaxation}
In rarer cases, a slower process is found,
designated as $\upalpha'$ and tentatively explained by
weak intermolecular associations.
While $\upgamma$ and higher order relaxations are always attributed to
innermolecular or side-chain motion,
the situation is less clear for $\upbeta$ relaxation.
There are theoretical arguments (Goldstein 1969) and 
experimental findings (Johari 1970) to support the belief
that $\upbeta$ relaxation is a universal property of
glass-forming systems.\index{Johari-Goldstein relaxation}
In any case, all secondary relaxations are well described
by the Arrhenius law~(\ref{EArrh}).
This implies that at some temperature they merge
with the $\upalpha$~relaxation.
This is well confirmed,
principally by dielectric spectroscopy,\index{dielectric spectroscopy}
which has the advantage of a particularly broad bandwidth.
Fig.~\ref{Fabrx} shows an example for this merger or decoupling
of $\upalpha$~and $\upbeta$~relaxation;
such an Arrhenius plot is also called
\D{relaxation map}\index{relaxation map}.

\subsection{The mode-coupling crossover}

\index{mode-coupling theory}

At present, there exists no satisfactory microscopic theory
of relaxation near the glass transition.
At moderate to low viscosities the situation is slightly better,
since basic phenomena can be explained to some extent by a
\D{mode-coupling theory} (G\"otze \etal\/ 1984--).\footnote
{The standard reference for mode-coupling theory
is the comprehensive book~\cite{Got09}.
More accessible introductions are provided by~\cite{BiKo05,Vog08}.}
This theory attacks the microscopic dynamics at the level
of the normalized density pair correlation function (\ref{EPhi}).
An equation of motion is written in the form
\begin{equation}\label{Emcteqmo}
  0 = \ddot\Phi_q(t) + \nu_q\dot\Phi_q(t) + \Omega_q^2\Phi_q(t) +
    \Omega_q^2 \int_0^t\!{\rm d}\tau\,\dot\Phi_q(t-\tau)m_q\{\Phi(\tau)\},
\end{equation}
which guarantees that subsequent approximations do not violate
conservation laws.
In a systematic, though uncontrolled expansion
the \D{memory kernel}\index{memory kernel} $m_q$
is then projected back onto products of pair correlations,
\begin{equation}
  m_q\{\Phi(t)\} \simeq \sum_{k+p=q} V_{kpq}(T) \Phi_k(t)\Phi_p(t).
\end{equation}
The coupling coeffients $V_{kpq}$
depend only on the static structure factor~$S(q)$,
which in turn depends weakly on the temperature.
This temperature dependence, however, is sufficient to trigger
a transition from ergodic, liquid-like solutions
\index{ergodicity}
to non-ergodic, glassy ones:
\index{T@$T_{\rm c}$}
\begin{equation}\label{Emctergo}
 \begin{array}{ll}
 \Phi_q(t\to\infty)\to0       &\mbox{ for }T>T_{\rm c},\\
 \Phi_q(t\to\infty)\to f_q>0  &\mbox{ for }T<T_{\rm c}.
 \end{array}
\end{equation}
On cooling towards $T_{\rm c}$,
a \D{critical slowing down}\index{critical slowing down} is predicted
that is characterized by \D{two} diverging timescales,
\begin{equation}\label{Emcttimes}
 \begin{array}{ll}
 t_\sigma=t_0\,\sigma^{-1/(2a)},\\
 \tau_\sigma=t_0\,\sigma^{-1/(2a)-1/(2b)},
 \end{array}
\end{equation}
with the reduced temperature $\sigma:=T/T_{\rm c}-1$.
The microscopic timescale $t_0$ is of the order $\Omega_q^{-1}$.
The exponents fulfill $0<a<b<1$; they depend on just one lineshape parameter
called $\lambda$.
The pair correlation function passes through the following
scaling\index{scaling} regimes:
\begin{equation}\label{Emctscaling}
 \begin{array}{ll}
  \Phi_q(t) \simeq f_q + h_q \sigma^{1/2} (t/t_\sigma)^{-a}
    &\mbox{ for }t_0\ll t\ll t_\sigma,\\
  \Phi_q(t) \simeq f_q - h_q \sigma^{1/2} B (t/t_\sigma)^{b}
    &\mbox{ for }t_\sigma\ll t\lesssim \tau_\sigma,\\
  \Phi_q(t) \simeq \hat\Phi_q(t/\tau_\sigma)
    &\mbox{ for }\tau_\sigma\lesssim t.\\
 \end{array}
\end{equation}
The regime delimited by the first two equations has been been given the 
unfortunate name ``fast $\upbeta$ relaxation''\index{$\upbeta$ relaxation}
(the Johari-Goldstein process\index{Johari-Goldstein relaxation}
 is then called ``slow $\upbeta$ relaxation''
though it is faster than $\upalpha$~relaxation).

The $t^{-a}$ power law is a genuine theoretical prediction;
it has been searched for in many scattering experiments,
and it has actually shown up in a number of liquids.
With the $t^b$ power law, the theory explains experimental facts
known since long (v.~Schweidler 1907).
This power law is also compatible with the short-time limit
of Kohlrausch's stretched exponential;\index{stretched exponential function}
it leads over to the $\upalpha$ relaxation master curve 
implied by the third equation of~(\ref{Emctscaling}).

\begin{figure} 
\centering
\includegraphics[width=0.48\textwidth]{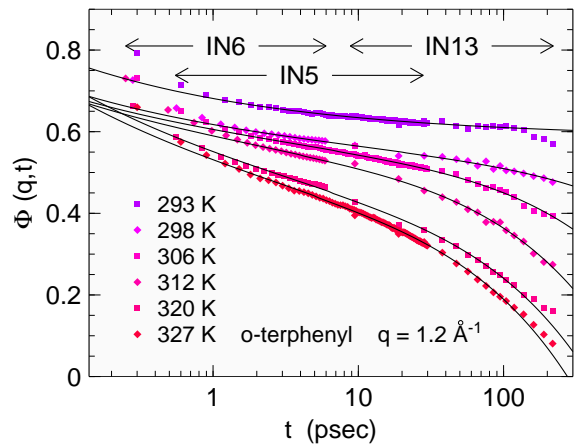}
\caption{Incoherent intermediate scattering function $S_{\rm self}(q,t)$
of the molecular glass former ortho-terphenyl,
obtained by combining Fourier transformed and resolution corrected
neutron scattering spectra from three different spectrometers.
Fits with the mode-coupling scaling law of fast $\upbeta$~relaxation.
Adapted from from~\cite{WuKB93}.}
\label{Fotp}
\end{figure} 

Mode-coupling predictions have been confirmed with impressive accuracy
by light scattering studies of the density driven glass transition
in colloidal suspensions (van Megen \etal\/ 1991--).\linebreak[3]
On the other hand, in conventional glass formers
mode coupling is not the full story.
By numerically solving simplified versions of (\ref{Emcteqmo}),
it is possible to fit relaxational spectra of normal or
slightly supercooled liquids.
On further supercooling,
in favorable cases (such as shown in Fig.~\ref{Fotp})
the power law asymptotes of slow $\upbeta$ relaxation appear,
and extrapolations yield a consistent estimate of $T_{\rm c}$.
Typically, this $T_{\rm c}$ is located 15\% to 20\% \D{above} $T_{\rm g}$.
This implies that the ergodicity breaking of Eq.~(\ref{Emctergo})
does not explain the glass transition;
nor does the power law (\ref{Emcttimes})
fit the divergence of viscosities or relaxation times near $T_{\rm g}$.

This leads to the view that $T_{\rm c}$ marks
a \D{crossover} between two dynamic regimes:
a mode-coupling regime at elevated temperature and low viscosity,
and a ``hopping'' regime in highly viscous, deeply supercooled liquids.
Experimental support for the significance of this crossover comes
from a possible change in the functional form of $\eta(T)$ or $\tau(T)$,
from the aforementioned breakdown of the
Stokes-Einstein relation%
\index{Stokes-Einstein relation}~(\ref{EEinsteinStokes}),
and from the merger of $\upalpha$~and slow $\upbeta$~relaxation.

\section{Dynamics of a Free Chain}

\subsection{The spring-bead model}\label{SKuhn}

\begin{figure} 
\centering
\includegraphics[width=0.45\textwidth]{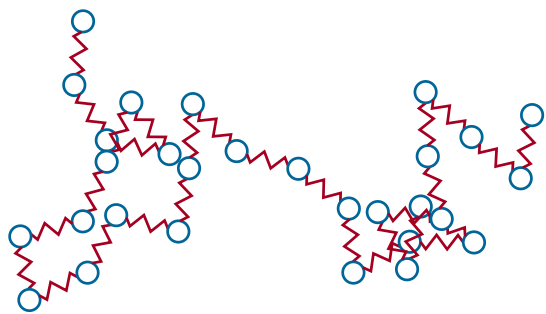}
\caption{Spring-bead model of a flexible polymer.
Since one bead represents several monomeric units,
there is no preferential bond angle at the level of this model.
Note also that the forces represented by the springs are entropic.}
\label{Fspring-bead}
\end{figure}

While the short-time dynamics of a polymer melt is quite similar 
to that of any other viscous liquid,
on longer time scales the chain structure of the polymer makes a
decisive difference, imposing strong constraints
on the segmental motion.
To study the motion of an entire chain,
we neglect all details of chemical structure.
We approximate the polymer
as a sequence of $N$ \textit{beads} 
at positions $\v{r}_n$ ($n=1,\ldots,N$),
connected by $N-1$
\textit{springs}.
Each bead represents several monomeric units
so that there is no preferred bond angle
\index{Kuhn model}
(Kuhn 1934[?], Fig.~\ref{Fspring-bead}).

The time-averaged equilibrium configuration of the polymer is assumed
to be given by a Gaussian distribution of bead-to-bead vectors
$\v{r}_n-\v{r}_{n-1}$,
\begin{equation}\label{EPGauss}
   P\{\v{r}\} \propto \exp\left(-\kappa\sum_{n=2}^{N}
     {\left(\v{r}_n-\v{r}_{n-1}\right)}^2 \right),
\end{equation}
where the force constant
\begin{equation}
  \kappa:=\frac{3k_{\rm B}T}{b^2}
\end{equation}
 ensures an average squared spring length of~$b^2$.

The size of a polymer coil\index{polymer radius} can be characterized by
a mean squared radius.
The most common measures are the
\D{end-to-end distance}\index{end-to-end distance} $R_{\rm e}$,
\begin{equation}\label{ERe}
  R_{\rm e}^2 := \langle {(\v{r}_N-\v{r}_1)}^2 \rangle
  = N b^2
\end{equation}
and the \D{gyration radius}\index{gyration radius}
\begin{equation}\label{ERg}
  R_{\rm g}^2 := N^{-1}\sum_{n}\langle {(\v{r}_n-\v{r}_{\rm G})}^2 \rangle
  \simeq N b^2 / 6,
\end{equation}
with the center of mass
\begin{equation}
  \v{r}_{\rm G}:=N^{-1}\sum\v{r}_n
\end{equation}
The expressions (\ref{ERe}), (\ref{ERg})
hold in polymer melts and in $\Theta$ solutions.
In other cases, both expressions are still valid approximations
if the factor $N$ is replaced by $N^{2\nu}$.
In good solvents, the exponent is $\nu=3/5$.

\subsection{The Rouse model}

The Gaussian distribution~(\ref{EPGauss}) is based on the assumption
that the free energy $A=U-TS$ is dominated
by the entropy $S=k_{\rm B}\ln P$ so that
the internal energy~$U$ can be neglected.
Hence each bead experiences an entropic force
\begin{equation}\label{Eentroforce}
   \v{F}_n^{\rm coil} = -\frac{\partial}{\partial \v{r}_n} A = 
     -\kappa\left(-\v{r}_{n-1} + 2\v{r}_n - \v{r}_{n+1}
                           \right),
\end{equation}
with obvious modifications for $n=1,N$.
This force strives to minimize the distances between beads,
thereby maximizing the coiling of the polymer.

The coupling to the heat bath shall be modelled by
 a random force $\v{F}_n^{\rm heat}$.
Its second moment is given by an obvious extension of~(\ref{Eramom2}),
\begin{equation}\label{Erandmom}
   \langle F^{\rm heat}_{n\alpha}(t) F^{\rm heat}_{m\beta}(t') \rangle
   = 2 \zeta k_{\rm B} T \delta_{nm} \delta_{\alpha\beta} \delta(t-t').
\end{equation}
Finally, moving beads experience a friction $-\zeta\partial_t\v{r}_n$.
These three forces make up the \D{Rouse model}\index{Rouse model},
which is a key reference in polymer physics (Rouse 1953).\footnote
{Rouse theory is covered in many text books,
most often using a continuum approximation
probably due to de Gennes \cite{Gen79}.
The short but well written chapter in \cite{BiKo05}
comes with a nice selection of experimental results.
Among dedicated polymer physics textbooks,
I found \cite{DoEd86} indispensable though largely indigestible
for its coverage of computational details,
and \cite{Str96} inspiring though sometimes suspicious
for its cursory outline of theory.}
Accordingly, the Langevin equation is
\begin{equation}\label{ELangevinRouse}
   m\partial_t^2\v{r}_n
   = -\zeta\partial_t\v{r}_n + \v{F}^{\rm coil}_n + \v{F}^{\rm heat}_n.
\end{equation}
In polymer solutions, the simple linear friction term is no longer adequate;
it must be replaced
by a \D{hydrodynamic interaction}\index{hydrodynamic interaction}.
This interaction is handled reasonably well by a theory
(Zimm 1956)\index{Zimm theory} outlined in Appendix~\ref{AZimm}.

At this point, let us indicate some orders of magnitude.
Assuming a spring length $b=1$~nm
and equating it with the hydrodynamic radius in~(\ref{EStokes}),
and assuming further a microscopic viscosity $\eta_{\rm s}\simeq10$~Pa$\cdot$s,
we find a friction coefficient\index{friction coefficient}~$\zeta$
 of the order of $10^{-7}$ to $10^{-6}$~Ns/m,
in agreement with empirical data
for polyisobutylene\index{polyisobutylene},
polymethyl acrylate\index{polymethyl acrylate}
and natural rubber\index{rubber}
\cite{Fer61}.
Assuming a bead mass of 100~Da,
the single-bead collision relaxation time~(\ref{Etaum})
is $\tau_m=m/\zeta\simeq 10^{-18}$~s,
which means that inertia is completely negligible on all relevant time scales.
At $T=300$~K, thermal motion is of the order
$\langle v^2\rangle^{1/2}\simeq 300$~m/s,
and the force constant is about $\kappa\simeq10^{-2}$~N/m.

\subsection{Rouse modes}

We will solve (\ref{ELangevinRouse}) by transforming to
normal coordinates.
To begin, we note that there is no coupling whatsoever between the three
Cartesian components.
Therefore, we need to consider just one of them.
Let us write $x$ and $f$ for an arbitrary component of $\v{r}$
and $\v{F}^{\rm heat}$. Then the Langevin equation reads
\begin{equation}\label{ELangevin2}
  m\partial_t^2 x_n 
  = -\zeta\partial_t x_n - \kappa \left( -x_{n-1} + 2x_n - x_{n+1} \right) + f_n.
\end{equation}
Introducing the vector notation $\V{x}:=(x_1,\ldots,x_{N})^{\rm T}$,
Eq.~(\ref{ELangevin2}) takes the form
\begin{equation}
   m\partial_t^2 \V{x} = -\zeta\partial_t \V{x} - \kappa \,\M{K}\,\V{x} + \V{f}
\end{equation}
with the force matrix
\begin{equation}
    \M{K} = \left( \begin{array}{cccccc}
            +1 & -1 & 0 & \cdots & 0 & 0 \\
            -1 & +2 & -1 & \ddots & 0 & 0 \\
            0 & -1  & +2 & \ddots & 0 & 0 \\
            \vdots & \ddots & \ddots & \ddots & -1 & 0 \\
            0 & 0 & 0 & -1 & +2 & -1 \\
            0 & 0 & 0 & 0 &  -1 & +1
           \end{array}\right)_{N\times N}.
\end{equation}
The entries $+1$ at both extremities of the diagonal
reflect the necessary modification of Eq.~(\ref{ELangevin2}) for $n=1,N$.
In the well known derivation of phonon dispersion,
this complication at the boundary is usually ignored
or superseded by an unphysical periodicity.
It is largely unknown that the correct $\M{K}$ can be diagonalized
quite easily without any approximation.
The eigenvalues are
\begin{equation}\label{Esin2}
    \lambda_p = 2 - 2 \cos\frac{p\pi}{N}
              = 4 \sin^2\frac{p\pi}{2N},\quad p=0,\ldots,N-1,
\end{equation}
and the normalized eigenvectors $\V{\hat{v}}_p$ have components
\begin{equation}\label{Evpn}
    \hat v_{pn} = \left\{
    \begin{array}{ll}
 \displaystyle      N^{-1/2} &\mbox{ for }p=0,\\
 \displaystyle      (N/2)^{-1/2}\cos\frac{p(n-\frac{1}{2})\pi}{N}
                         &\mbox{ for all other $p$},
    \end{array} \right.
\end{equation}
where $n=1,\ldots,N$.
The proof requires no more than a straightforward verification of
$\M{K}\,\V{\hat v}_p=\lambda_p\V{\hat v}$.
Collecting the normalized eigenvectors into an orthogonal matrix
$\M{A}:=(\V{\hat v}_0,\ldots,\V{\hat v}_{N-1})$,
we introduce normal coordinates
\begin{equation}\label{xAx}
  \V{\tilde x}:= \M{A}^{\rm T} \V{x},\quad
  \V{x} = \M{A} \,\V{\tilde x},
\end{equation}
and similar for $\V{f}$.
It is easily seen that the average random force correlation (\ref{Erandmom})
is still diagonal,
\begin{equation}\label{Erandmom2}
   \langle \tilde{f}_p(t) \tilde{f}_q(t') \rangle
   = 2 \zeta k_{\rm B} T \delta_{pq}  \delta(t-t').
\end{equation}
In consequence, for each normal mode one obtains a decoupled
Langevin equation
\begin{equation}\label{Erouselangevin8}
   m\partial_t^2 \tilde{x}_p
   = -\zeta\partial_t \tilde{x}_p - \kappa\lambda_p \tilde{x}_p + \tilde{f}_p.
\end{equation}
At this point we must distinguish the eigenmode with the special
eigenvalue $\lambda_0=0$ from all the others.

The eigenmode $p=0$ describes the
motion of the center of mass~$\v{r}_{\rm G}=N^{-1/2}\v{\tilde r}_0$.  
Since $\lambda_0=0$, (\ref{Erouselangevin8}) is identical with
the Langevin equation for Brownian motion (\ref{ELv1}).
Accordingly, the long-time evolution of the
mean squared displacement\index{mean squared displacement}
is $\langle r_{\rm G}^2(t)\rangle\simeq 6D_{\rm R}t$,
where the macromolecular
\D{Rouse diffusion coefficient}\index{diffusion coefficient}
is given by a rescaled version
of the Sutherland-Einstein relation~(\ref{EEinstein}):
\begin{equation}
  D_{\rm R}=\frac{k_{\rm B}T}{\zeta N}.
\end{equation}

Turning to the \D{Rouse modes}\index{Rouse mode} with $p>0$,
we neglect the inertial term in the Langevin equation~(\ref{Erouselangevin8}).
Integration is then straightforward:
\begin{equation}\label{Etxpt}
  \tilde{x}_p(t)
  = \zeta^{-1}\int_{-\infty}^t\!{\rm d}t'\,{\rm e}^{-(t-t')/\tau_p}\tilde{f}_p(t'),
\end{equation}
introducing the \D{mode relaxation time} $\tau_p:=\zeta/(\kappa\lambda_p)$.
It can be approximated for $p\ll N$ as
\begin{equation}\label{Etaup}
    \tau_p  \simeq \frac{\tau_{\rm R}}{p^2}.
\end{equation}
The relaxation time of the fundamental mode $p=1$
is known as the \D{Rouse time}\index{Rouse time}:
\begin{equation}\label{EtauR}
  \tau_{\rm R} := \frac{L^2\zeta}{3\pi^2 k_{\rm B}T }.
\end{equation}
In this expression, $N$ and $b$ enter only
via the \D{extended chain length}\index{extended chain length} $L=Nb$,
which does not change if we change our spring-bead model to consist
of $N/x$ segments of length $bx$.
This justifies \D{ex post} that
we have ignored all details of microscopic structure.
A simple alternative expression for $\tau_{\rm R}$ is
\begin{equation}\label{EtauR2}
  \tau_{\rm R} := \frac{2 R_{\rm g}^2}{\pi^2 D_{\rm R}}
\end{equation}
with the gyration radius~(\ref{ERg}).

\subsection{Macroscopic consequences: dielectric and shear response}

\begin{figure} 
  \begin{minipage}[b]{0.25\textwidth}
\includegraphics[width=1\textwidth]{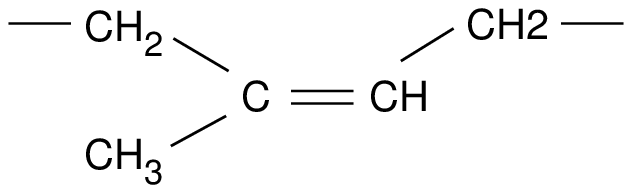}
  \vspace{6ex}\strut  
  \end{minipage}
\hfill
\includegraphics[width=0.42\textwidth]{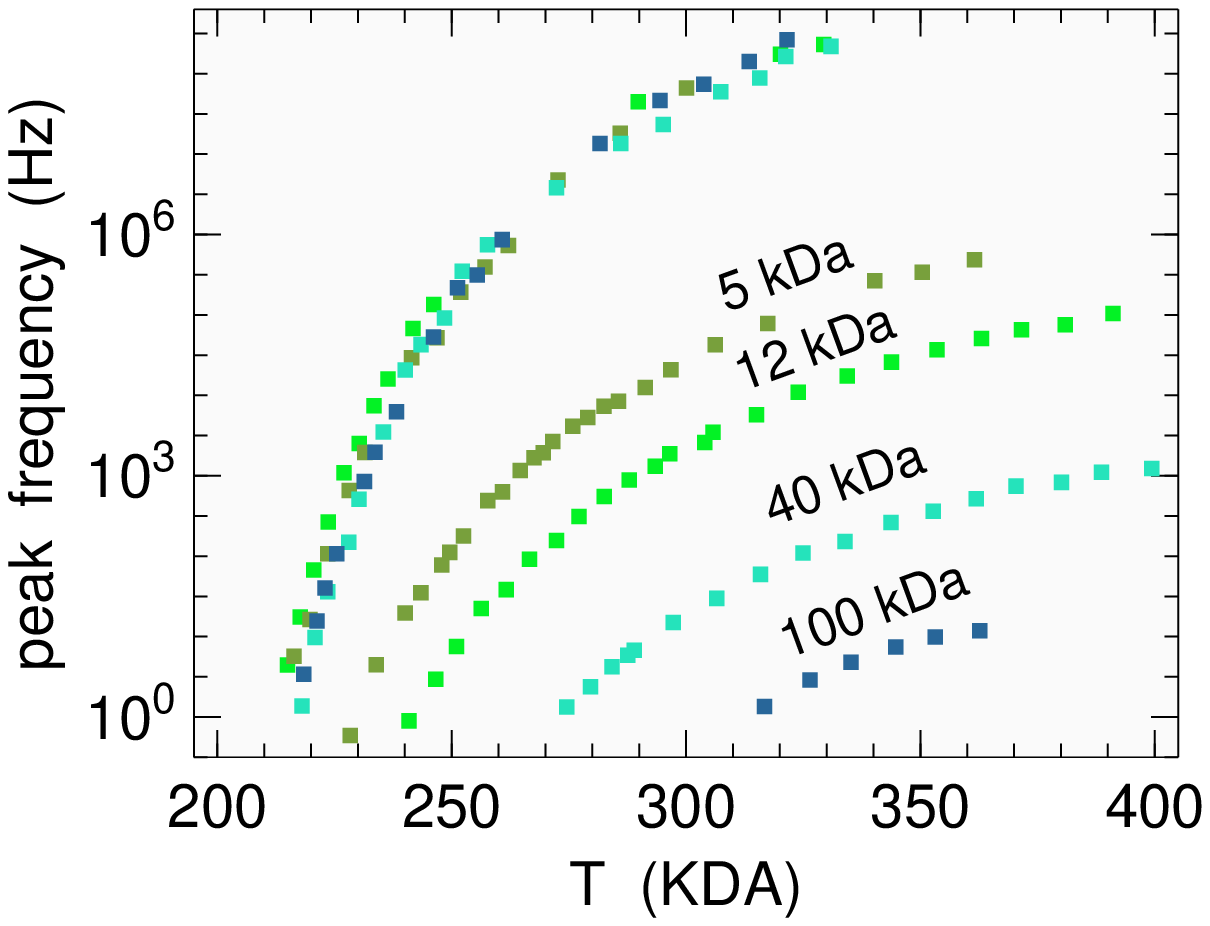}
\hfill
\includegraphics[width=0.27\textwidth]{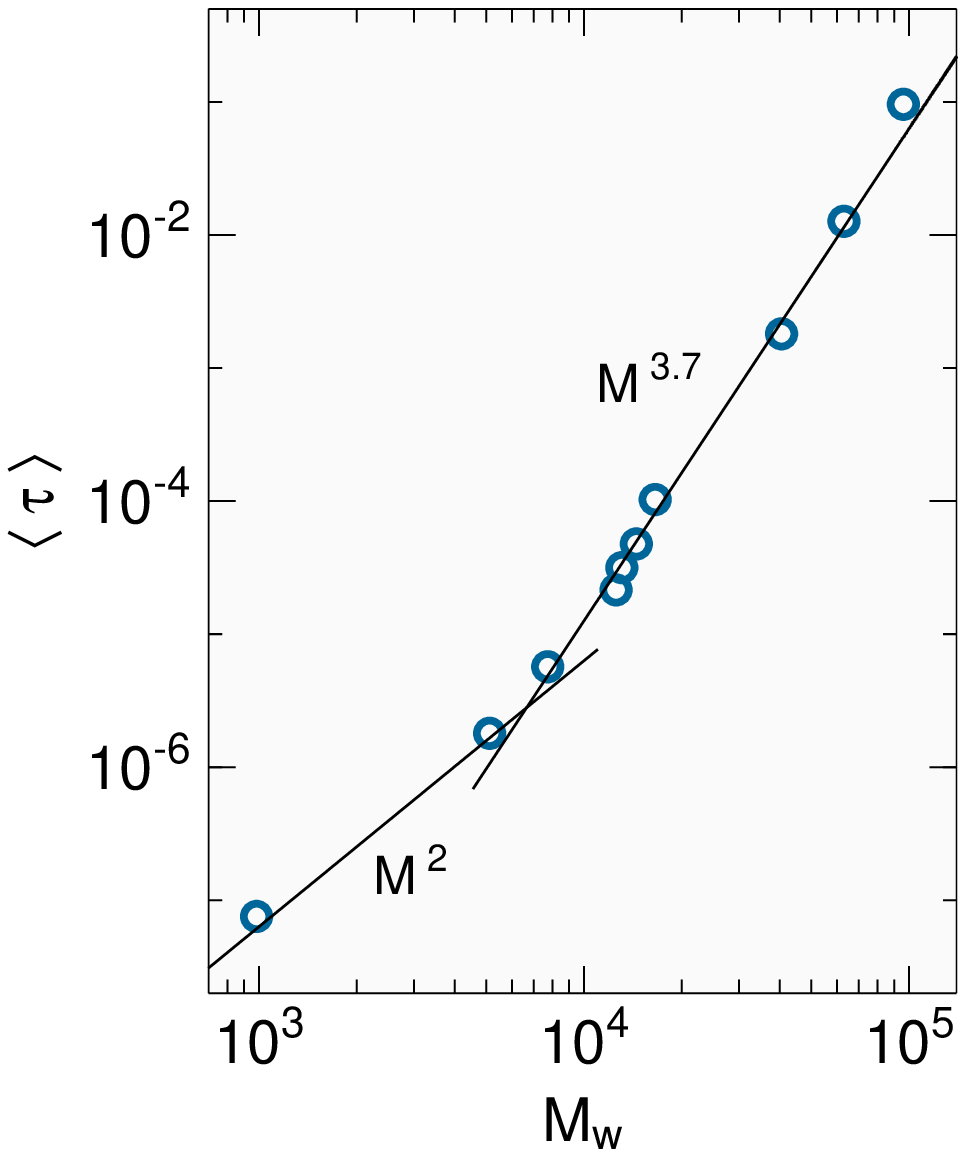}
\caption{(a) Cis-polyisoprene consists of monomers
that possess a dipole moment in direction of the polymeric bonds.
(b) Peak frequencies $\omega$ of dielectric loss spectra
in cis-polyisoprene samples of different molecular weight~$M$.
The normal mode resonance depends strongly on $M$,
whereas the $\alpha$~relaxation does not.
(c) Normal mode relaxation times $\tau=\omega^{-1}$ versus~$M$,
showing a crossover from the Rouse regime $\tau_{\rm R}\sim M^2$ to
a stronger $M$ dependence above the entanglement onset $M_c$.
Data from \cite{BoKr90}.}
\label{FcisPI}
\end{figure} 

In favorable cases, Rouse relaxation can be observed 
by dielectric spectroscopy.
In the simplest case, the monomeric possess a dielectric moment~$\mu$
in direction of the polymeric bond.
This requires the absence of a perpendicular mirror plane,
which is the case e.~g.\ for
cis-polyisoprene\index{cis-polyisoprene|see{polyisoprene}}\index{polyisoprene}
(Fig.~\ref{FcisPI}a).
Then the overall dipole moment of the macromolecule is
\begin{equation}\label{Emu}
  \v{\mu} = \sum_{n=2}^{N} (\v{r}_n - \v{r}_{n-1}) \mu 
  = (\v{r}_{N} - \v{r}_1) \mu
\end{equation}
From (\ref{Evpn}) we infer that
only eigenmodes with odd~$p$ contribute to (\ref{Emu}).
For $t\gtrsim\tau_{\rm R}$ pair correlations are dominated by the $p=1$ mode:
\begin{equation}
  \langle\v{\mu}(0)\v{\mu}(t)\rangle
  \propto \langle\v{\tilde{r}}_1(0)\v{\tilde{r}}_1(t)\rangle
  \propto {\rm e}^{-t/\tau_{\rm R}}.
\end{equation}
According to the fluctuation-dissipation theorem (\ref{EFDT}),
this correlation is proportional to a linear response function.
After Fourier transform,
one finds that the dielectric permittivity $\epsilon(\omega)$
has a Debye resonance around $\omega\sim\tau_{\rm R}^{-1}$.

Fig.~\ref{FcisPI}b,c shows results 
of dielectric spectroscopy in cis-polyisoprene melts
with different extended chain lengths $L=Nb$.
In experimental reports the chain length is of course expressed
as molecular weight~$M$.
The $\alpha$~relaxation peak,
discussed above in Sect.~\ref{Sarx},
is perfectly independent of~$M$,
which confirms that it is due to innersegmental motion,
In contrast, the normal mode relaxation time depends strongly on~$M$.
For low~$M$, the Rouse prediction $\tau_R\sim M^2$ is confirmed.
However, at $M_c \simeq 10$~kDalton, there is a rather sharp crossover
to a steeper power law $M^{3.7}$ that is ascribed to entanglement effects.
\index{entanglement}

Mechanical spectroscopy has the advantage that it works also
if monomers are too symmetric for dielectric measurements.
As already mentioned in Sect.~\ref{Sarx}
\D{torsional spectroscopy}\index{torsional spectroscopy}
probes a system's response to \D{shear}.
Not seldom this response is nonlinear,
due to non-Newtonian flow\index{non-Newtonian flow} phenomena that are beyond
the scope of the present lecture.
As long as the response is linear,
it can be described by the frequency dependent
\D{shear modulus}\index{shear modulus} $G(\omega)$.
Its high frequency limit $G_\infty$ quantifies the
shear needed to cause a given \D{strain}.
In a \D{liquid}, the low frequency limit $G_0$ is zero:
stationary shear is not able to build up a lasting stress.
Sometimes this is seen as the defining property of the liquid state.
Instead of stationary shear, a flow gradient is needed to maintain strain.
The proportionality coefficient is the
\D{shear viscosity}\index{shear viscosity}\index{viscosity|see{shear
viscosity}}~$\eta$, which is the low frequency limit of $G(\omega)/(i\omega)$,
as anticipated in~(\ref{EGliq}).

To calculate $G(\omega)$ in the frame of Rouse theory,
it is convenient to invoke
a \D{Green-Kubo relation}\index{Green-Kubo relation}
(a variant of the fluctuation-dissipation theorem)
according to which $G(\omega)$ is proportional to 
the Fourier transform of a stress autocorrelation function 
$\langle\sigma_{xz}(t)\sigma_{xz}(0)\rangle$.\footnote
{The following is no more than a speculative summary of obscure calculations
in \cite{DoEd86} and \cite{Str96}.}
The stress component $\sigma_{xz}$ can be expressed through the 
displacements of individual beads.
For weak elongations,
the autocorrelation can be factorized so that
$G(t)$ is proportional to the square
of a normalized one-dimensional displacement
autocorrelation function
\begin{equation}
  \sum_{n} \frac{\langle \delta x_n(t)\delta x_n(0)\rangle}
                {\langle \delta x_n^2\rangle}
  = \sum_{p\ge1} \frac{\langle\tilde{x}_p(t)\tilde{x}_p(0)\rangle}
                     {\langle\tilde{x}_p^2\rangle}.
\end{equation}
Using (\ref{Etxpt}) to compute the normal mode autocorrelation we find
\begin{equation}\label{EGSum}
  G(t) \sim \sum_p {\rm e}^{-2t/\tau_p}.
\end{equation}
With $\tau_p\propto p^{-2}$ and replacing the sum by an integral
we obtain the approximation
\begin{equation}\label{EGGauss}
  G(t) \sim \sum_{p\ge1} {\rm e}^{-2p^2t/\tau_{\rm R}}
  \sim \int_0^\infty\!{\rm d}p\,{\rm e}^{-2p^2t/\tau_{\rm R}}
  \sim {\left(\frac{\tau_{\rm R}}{t}\right)}^{1/2}.
\end{equation}
Fourier transform yields the power law
\begin{equation}\label{EGsqrtw}
  G(\omega) \sim \omega^{1/2}.
\end{equation}
For small~$t$, quite many eigenmodes contribute to (\ref{EGSum})
so that neither the low-$p$ expansion~(\ref{Etaup}) nor
the extension of the summation to~$\infty$ are justified.
Therefore $G(\omega)$ must cross over from (\ref{EGsqrtw})
to a constant high-frequency limit~$G_\infty$.

\begin{figure} 
\centering\includegraphics[width=0.5\textwidth]{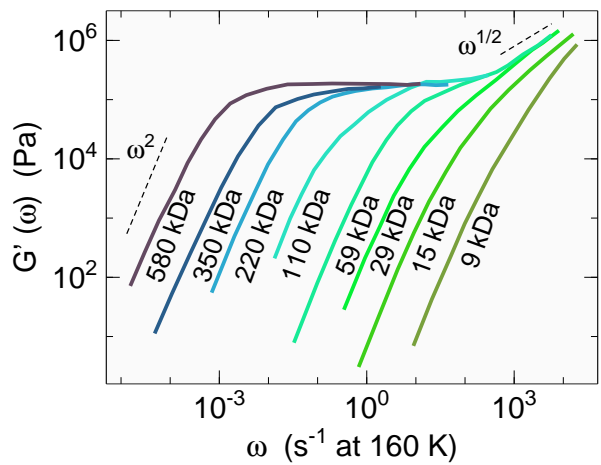}
\caption{Real part $G'(\omega)$ of the shear modulus of polystyrene samples
with different narrow molecular weight distributions.
Experiments were done in a frequency range
from $10^{-1.5}$ to $10^{0.5}$~s$^{-1}$ and
in a temperature range from 120 to 260~K.
Then, time-temperature superposition was used to construct the
master curves shown here. Data from~\cite{OnMK70}.}
\label{FGw}
\end{figure} 

For large $t$, only few eigenmodes contribute to (\ref{EGSum})
so that the passage to a continuous~$p$ becomes invalid.
In this case, it is more appropriate to compute 
the Fourier transform term by term, which yields a sum
of Maxwell-Debye resonances
\begin{equation}\label{EGMaxw}
  G(\omega) \sim \sum_p \frac{1}{1-i\omega \tau_p}.
\end{equation}
In the low-frequency limit we obtain a constant term of doubtful significance
plus a linear term proportional to the viscosity\index{shear viscosity}
\begin{equation}
  \eta\sim\sum\tau_p\sim\tau_{\rm R}.
\end{equation}
From (\ref{EtauR}), we have $\tau_{\rm R}\sim N^2$,
but there is a prefactor $N^{-1}$ in $G(\omega)$ omitted in our sloppy
derivation so that finally the Rouse model predicts $\eta\sim N$.

In Fig.~\ref{FGw} experimental data are shown.
For moderate chain lengths,
$G'(\omega)$ is indeed found to cross over from~$\omega^2$
(the lowest non-constant real term in the expansion of (\ref{EGMaxw})
to $\omega^{1/2}$.
The ultimate limit $G_\infty$ has not been reached in this experiment.
For longer polymer chains,
a flat plateau appears between 
the liquid-like~$\omega^2$ and the Rouse regime~$\omega^{1/2}$.
Such a constant value of $G(\omega)$ implies instantaneous,
memory-free response,
which is characteristic of rubber elasticity;\index{rubber}
it is caused by entanglement.\index{entanglement}

\subsection{Microscopic verification: neutron spin echo}

For direct, microscopic measurement of chain conformation fluctuations
one must access length and time scales of the order of nm and~ns.
The most powerful instrument in this domain is the 
\D{neutron spin echo}\index{neutron spin echo} spectrometer.
The recent book~\cite{RiMA05} provides a comprehensive review of
spin echo studies on polymer dynamics.

Usually, spin echo experiments require deuterated samples
to avoid the otherwise dominant incoherent scattering from protons.
The experiments then yield the
\D{coherent dynamic structure factor}\index{coherent
  neutron scattering}~$S(q,t)$.
However, for a simple, intuitive data analysis 
the \D{incoherent} scattering function\index{incoherent
  neutron scattering}~$S_{\rm i}(q,t)$ is preferable.
It is also denoted $S_{\rm self}(q,t)$ since it reveals
the \D{self correlation}\index{self correlation}
of a \D{tagged particle}\index{tagged particle}.
In \D{Gaussian approximation}\index{Gaussian approximation}
 \begin{equation}\label{Siqt}
   S_{\rm i}(q,t)=\langle \exp\left( iq(\v{r}(t)-\v{r}(0))\right) \rangle
    \simeq \exp\left( -q^2\langle r^2(t)\rangle/6\right),
 \end{equation}
it yields the
{mean squared displacement}\index{mean squared displacement}~(\ref{Er2t}).

Measuring~$S_{\rm i}(q,t)$ by neutron spin echo is difficult
because the random spin flips associated with
incoherent scattering destroy 2/3 of the incoming polarization.
Nevertheless, thanks to progress in instrumentation,
it is nowadays possible to obtain self correlation functions 
from undeuterated (``protonated'') samples in decent quality.
Alternatively, 
self correlations can be measured 
with the data quality of coherent scattering if short protonated
sequences are intercalated at random in deuterated chains.

Within the Rouse model,
and neglecting ballistic short-time terms,
the mean squared displacement is given by
\begin{equation}
  \langle r^2(t)\rangle
  = \frac{6}{N}\sum_{p=0}^{N-1}\left[
           \langle \v{\tilde x}_p^2\rangle 
        -  \langle \v{\tilde x}_p(t)\v{\tilde x}_p(0)\rangle \right]
  = 6 D_{\rm R} \left\{ t + \sum_{p=1}^{N-1} \tau_p\,
          \left[1-{\rm e}^{-t/\tau_p}\right] \right\}.
\end{equation}
With the same technique as in~(\ref{EGGauss}),
one obtains the approximation
\begin{equation}\label{Er2tRouse}
  \langle r^2(t)\rangle
  \simeq 6 D_{\rm R} \left\{ t + {\left(\pi\tau_{\rm R}t\right)}^{1/2} \right\}.
\end{equation}
At about $t \sim \tau_{\rm R}$,
there is a cross-over from a $t^{1/2}$ regime dominated
by conformational fluctuations to the $t^1$ diffusion limit.
Inserting the asymptotic $\langle r^2(t)\rangle\sim t^{1/2}$
into~(\ref{Siqt})
one obtains an expression that agrees
with the
Kohlrausch function~(\ref{EKohl})\index{stretched exponential function}
with a stretching exponent $\beta=1/2$.
This indicates that the high-$p$ limit of the Rouse modes
is more physical than might have been expected;
it seems to capture even some aspects of segmental $\alpha$ relaxation.
Be that as it may,
the $t^{1/2}$ prediction has been impressively confirmed in
neutron scattering experiments.

\begin{figure} 
\strut
\hfill
\includegraphics[width=0.94\textwidth]{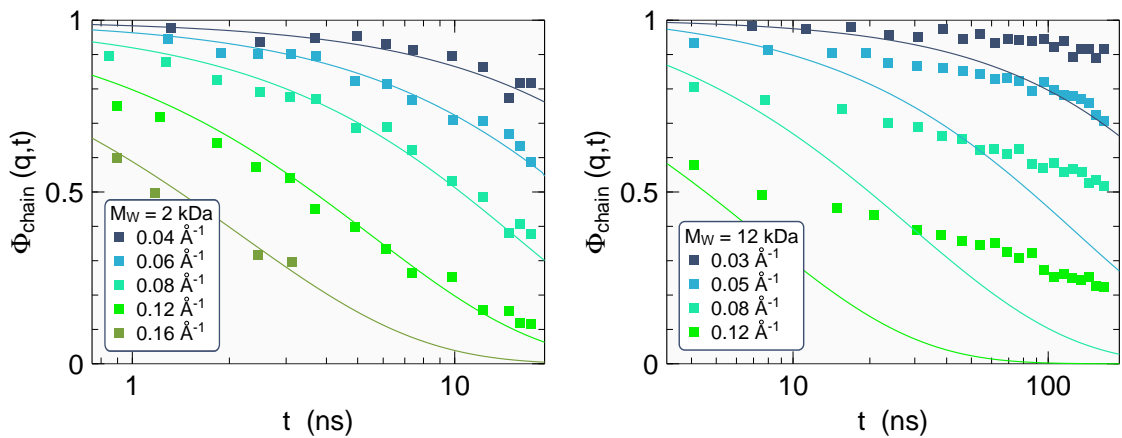}
\hfill
\strut
\caption{Single-chain coherent normalized dynamic structure factor
of polyethylene\index{polyethylene} melts at 509~K.
Lines are predictions of the Rouse model.
They fit for short chains (left), but not for long chains (right);
note the different time scales.
Data points and lines from~\cite{RiMA05}.}
\label{F315}
\end{figure} 

The \D{coherent} dynamic structure factor is more involved than~(\ref{Siqt}).
In general, it contains contributions
from interchain as well as from intrachain correlations.
Single-chain dynamics can be isolated
by \D{contrast variation}\index{contrast variation},
using a mixture of about 10\% protonated and 90\% deuterated polymer.
Fig.~\ref{F315} shows the single-chain dynamic structure factor
of two polyethylene melts with different chain lengths~\cite{RiMA05}.
For short chains, the data are in perfect agreement with the Rouse model.
For long chains, however,
correlations decay much slower than predicted by the Rouse model.
This is yet another indication of entanglement.

\section{Entanglement and Reptation}

\begin{figure} 
\strut
\hfill
\includegraphics[width=0.32\textwidth]{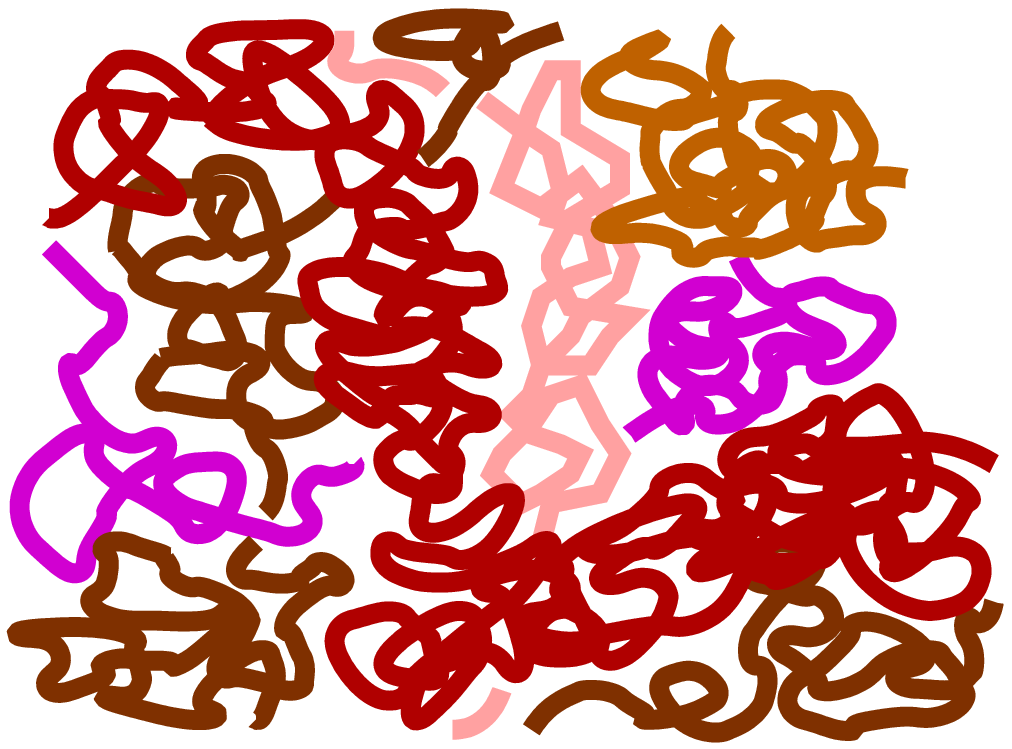}
\hfill
\includegraphics[width=0.38\textwidth]{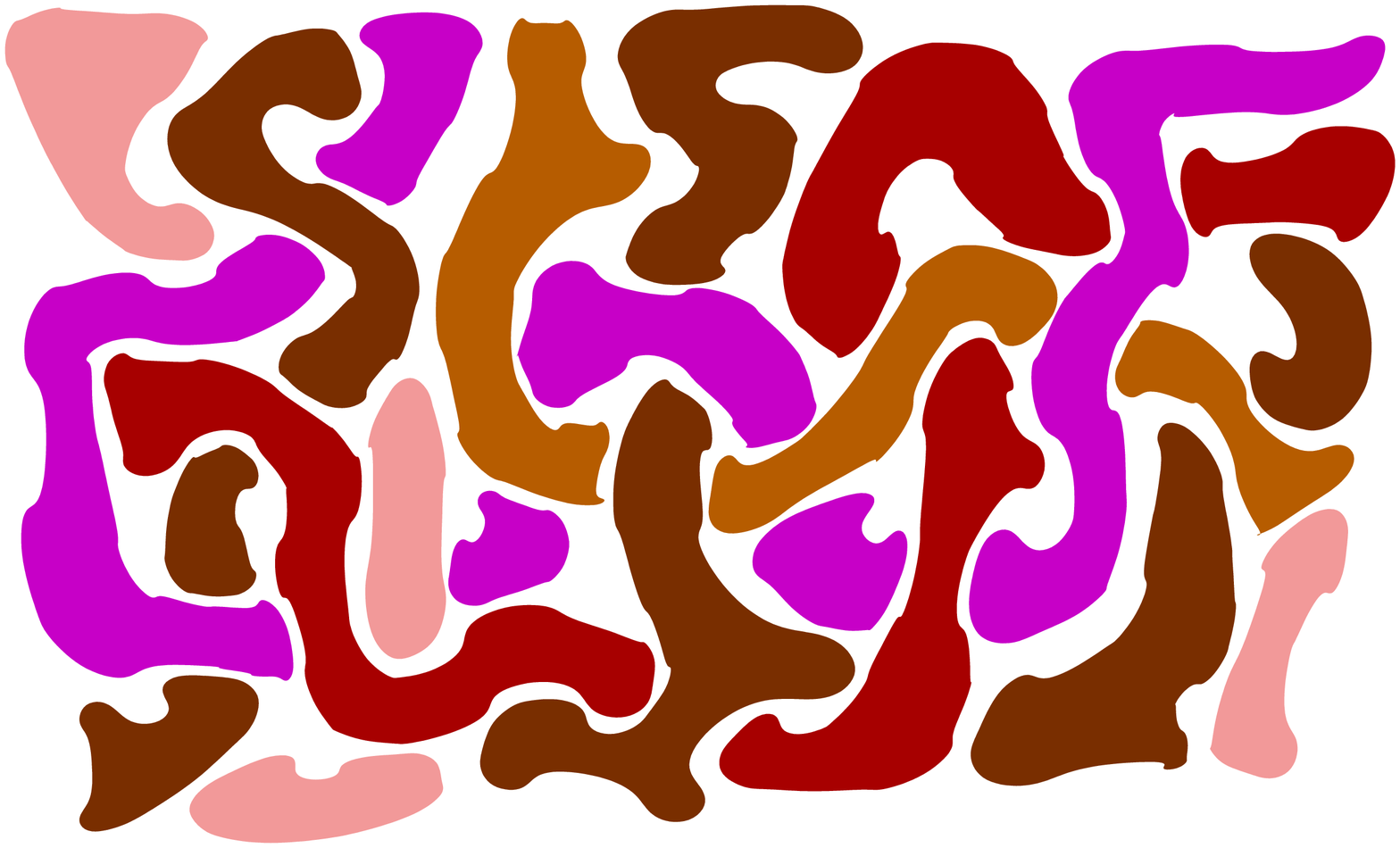}
\hfill
\strut
\caption{Entanglement.
(a) Local detail, with relatively few chain crossings.
(b) Coarse grained representation, 
showing heavy entanglement.
Each monochrome, tube-like region is
filled with coiled subunits of the polymer.
For times between the entanglement time $t_{\rm e}$ and
the disentanglement time~$t_{\rm d}$,
polymer motion is confined to these tubes.}
\label{Fentangle}
\end{figure}

Entanglement\index{entanglement}
 means that the conformational dynamics of a chain
is hindered by presence of other chains.
Entanglement is a topological constraint,
due to the simple fact that chains cannot cross each other
(Fig.~\ref{Fentangle}).
We have already encountered experimental results
that provide clear evidence for the relevance of entanglement
for polymer chains that exceed a certain size~$N_{\rm c}$:
\begin{itemize}
\item The dielectric relaxation time crosses over
from the Rouse behavior $\tau\propto N^2$ (\ref{EtauR})
to a steeper slope  $\tau\propto N^{3.7}$ (Fig.~\ref{FcisPI}c).
\item In the shear modulus $G(\omega)$,
there appears a plateau $G(\omega)=\mbox{const}$ with rubber-like elasticity
between the liquid limit $G(\omega)\simeq i\eta\omega$ and
the Rouse regime $G(\omega)\sim\omega^{1/2}$ (Fig.~\ref{FGw}).
\item Neutron scattering shows that correlations within long chains
decay much slower than predicted by the Rouse model~(Fig.~\ref{F315}).
\end{itemize}
The rather sharp crossover at~$N_{\rm c}$
implies that entanglement becomes relevant if the coil radius~$N^{1/2}b$
(up to a constant factor, depending on definition)
exceeds a certain value
\begin{equation}
  a:=N_{\rm c}^{1/2}b.
\end{equation}
Up to this length scale, the chains are heavily coiled with little
mutual penetration (Fig.~\ref{Fentangle}a).
On larger scales,
the coarse-grained polymer chains have the character of
heavily entangled \D{tubes}\index{tube model} (Fig.~\ref{Fentangle}b).
Each tube can be modelled as an ideal random chain,
consisting of beads of size~$a$.

An \D{entanglement time}\index{entanglement time}~$t_{\rm e}$
can be defined by $\langle r^2(t_{\rm e})\rangle\simeq a^2$.
Using~(\ref{Er2tRouse}) in the limit $t\ll\tau_{\rm R}$,
we find up to numeric prefactors
\begin{equation}
  t_{\rm e}\sim \frac{L_{\rm c}^2}{D}
\end{equation}
with the critical extended chain length\index{extended chain length}
 $L_{\rm c}=N_{\rm c}b$.
For times beyond $t_{\rm e}$,
the dynamics is qualitatively different 
from the free chain Rouse regime.
Since a chain is basically confined to a tube,
it can only perform a one-dimensional, snake-like motion,
called \D{reptation}\index{reptation} (de Gennes 1971).

\begin{figure} 
\strut
\hfill
\includegraphics[width=0.43\textwidth]{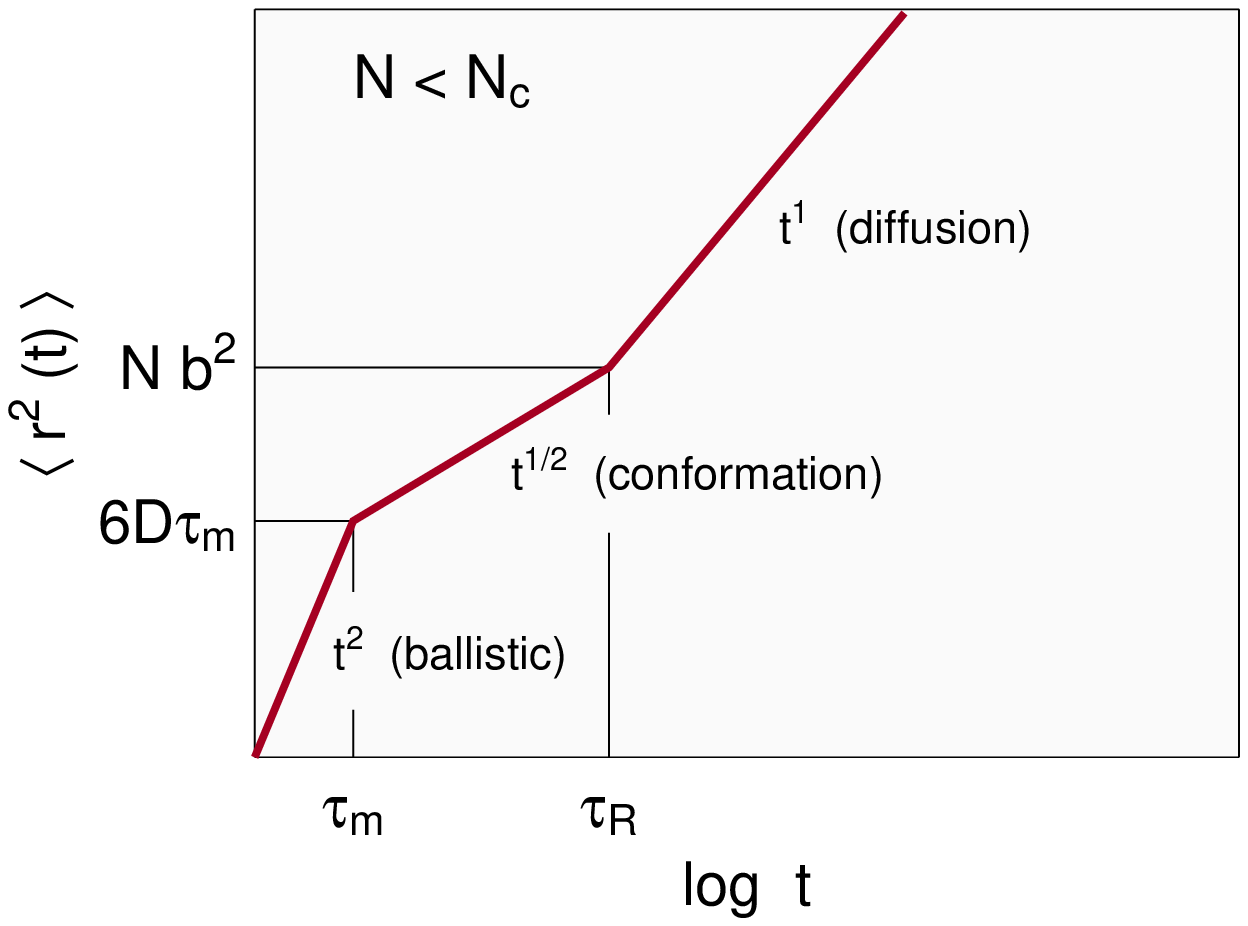}
\hfill
\includegraphics[width=0.43\textwidth]{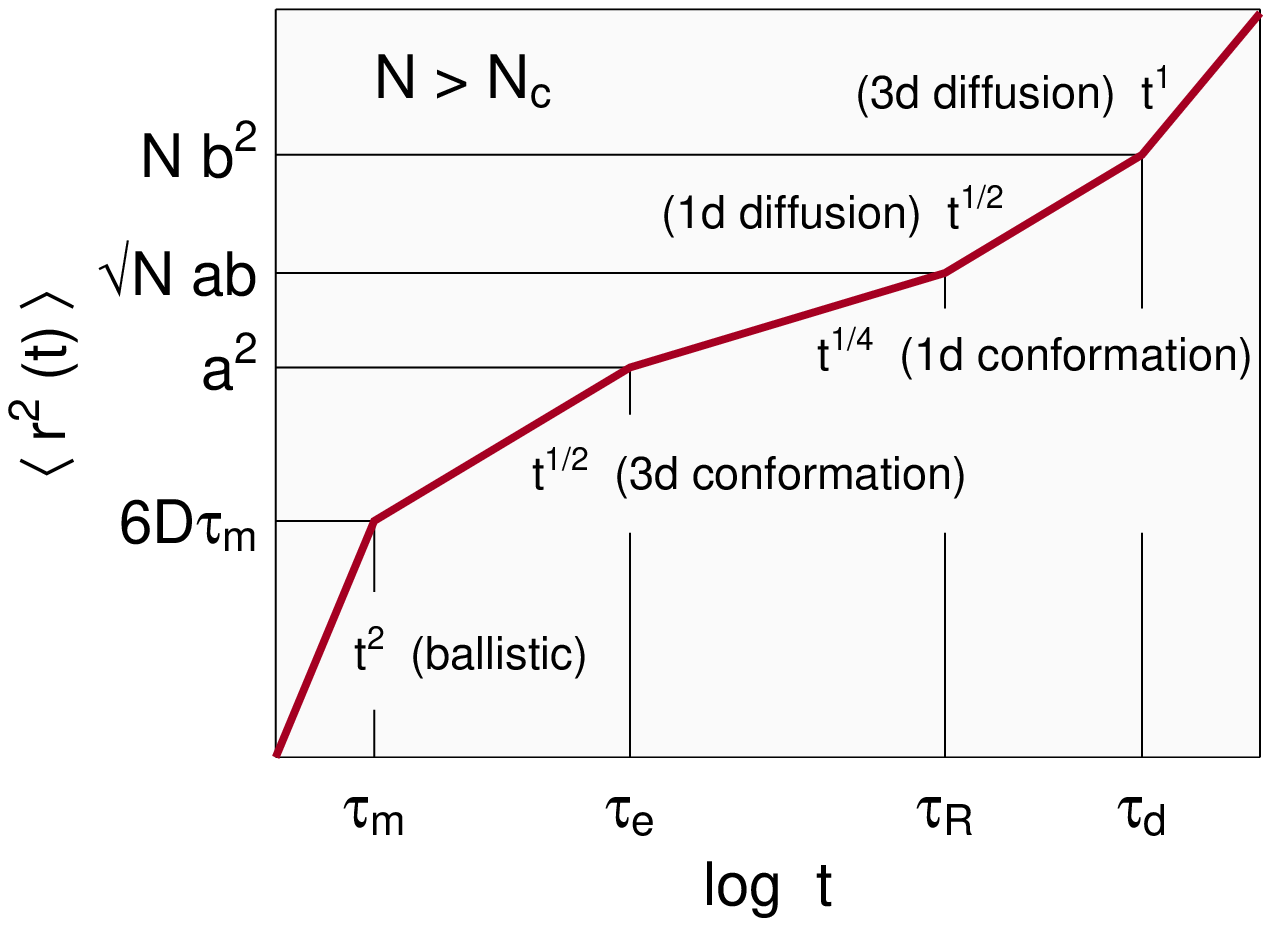}
\hfill
\strut
\caption{Time evolution of the mean squared displacement (\ref{Er2t})
on a double logarithmic scale.
(a) For short chains, as predicted by Rouse theory.
(b) For long chains, as predicted by de Gennes' reptation theory.}
\label{Fr2t}
\end{figure}

For a short outline of some scaling\index{scaling} results,
we concentrate on the
mean squared displacement\index{mean squared displacement}.
We will see that there are altogether
no less than five different regimes.
They are summarized in Fig.~\ref{Fr2t}.

The one-dimensional dynamics within a tube shall be
described by the Rouse model as before.
Let~$s$ be a coordinate along the tube.
The mean squared displacement in~$s$
is just one third of the Rouse result~(\ref{Er2tRouse}) in~$\v{r}$.
Since the tubes are ideal Gaussian random coils,
an extended tube length $s^2=N_s^2 a^2$ corresponds 
to a squared real-space displacement of $r^2=N_s a^2=as$.
In the \D{reptation} regime $t_{\rm e}\ll t\ll t_{\rm R}$
we obtain, omitting prefactors,
\begin{equation}\label{Er2tRep}
  \langle r^2(t)\rangle
  \sim a D_{\rm R}^{1/2}{(\tau_R t)}^{1/4}.
\end{equation}
This $t^{1/4}$ law is a key prediction of reptation theory;
it has been confirmed by neutron spin echo measurements \cite{RiMA05}.

For times beyond $t_{\rm R}$,
the one-dimensional dynamics crosses over
from innerchain Rouse fluctuations to center-of-mass diffusion.
Accordingly, the real-space mean squared displacement
takes the form
\begin{equation}\label{Er2t1dif}
  \langle r^2(t)\rangle
  \sim a D_{\rm R}^{1/2}t^{1/2}.
\end{equation}
This holds until
the chain escapes from its tube,
which happens when $s^2\sim N^2 b^2$ or $r^2\sim a D_{\rm R}t_{\rm d}^{1/2}$.
Using again $r^2=as$,
we obtain the
\D{disentanglement time}\index{disentanglement
  time}
\begin{equation}
  t_{\rm d}\sim \frac{N^3 b^2}{D}.
\end{equation}

Finally, on time scales above $t_{\rm d}$,
the chain, having diffused out of its original tube,
is free to try new conformations in three dimensions.
This is a center-of-mass random walk, described by
\begin{equation}\label{Er2t3dif}
  \langle r^2(t)\rangle
  \simeq 6 D_{\rm d} t.
\end{equation}
Matching (\ref{Er2t3dif}) with~(\ref{Er2t1dif})
at $t_{\rm d}$, we obtain the
\D{disentangled diffusion coefficient}\index{diffusion coefficient}
\begin{equation}
  D_{\rm d} \sim \frac{aD}{b N^2}.
\end{equation}

The scaling laws $t_{\rm d}\sim N^3$ and $D_{\rm d} \sim N^{-2}$
are important predictions.
The disentanglement time $t_{\rm d}$ determines
the relaxation time~$\tau$ observed in dielectric or mechanical spectroscopy.
Empirically, the molecular mass dependence of~$\tau$
is even stronger than~$N^3$.
Typical exponents are 3.2 to 3.6;
in Fig.~\ref{FcisPI}c we had even 3.7.
This discrepancy shows that one-dimensional diffusion in fixed tubes
is not the full story.
It is necessary to take into account fluctuations of the neighbouring chains
(\D{contour length fluctuations}\index{contour length fluctuation}).
On the other hand, the prediction $D_{\rm d} \sim N^{-2}$
has been confirmed by quite direct, spatially resolved 
diffusion measurements~\cite{Str96}.

\section*{Appendices}

\appendix

\section{Linear response theory: relaxation, dissipation, fluctuation}\label{ALinResp}

If a multi-particle system is exposed to a weak perturbation $A$,
its response $B$ is linear in $A$, as far as amplitudes are concerned.
However, the response may be delayed in time, assuming the character of
\D{relaxation}\index{relaxation}.
Relaxation may be probed in time or in frequency,
by spectroscopy (response to external perturbation)
or by scattering methods (fluctuations in equilibrium).
The relations between these probes 
are the subject of \D{linear response theory}\index{linear response},
to be briefly summarized in this appendix.

The linear response $B(t)$ to a perturbation $A(t)$ can be written as
\begin{equation}\label{EALR}
  B(t) = \int_{-\infty}^t\!{\rm d}t'\,R(t-t')\,A(t').
\end{equation}
Consider first the momentary perturbation $A(t)=\delta(t)$.
The response is $B(t)=R(t)$. Therefore, the memory kernel $R$ is
identified as the \D{response function}.

Consider next a perturbation $A(t)={\rm e}^{\eta t}\Theta(-t)$
that is slowly switched on and suddenly switched off
($\Theta$ is the Heavyside step function,
$\eta$ is sent to $0^+$ at the end of the calculation).
For $t>0$, one obtains $B(t)=\Phi(t)$
where $\Phi$ is the negative primitive of the response function,
\begin{equation}\label{ERPhi}
  R(t)=-\partial_t\Phi(t).  
\end{equation}
Since $\Phi$ describes the time evolution after an external perturbation
has been switched off, it is called the \D{relaxation function}.
In the special case of exponential
 (Lorentzian)\index{Lorentzian|see
 {exponential relaxation}}\index{exponential relaxation}
relaxation, $R$ and $\Phi$ are equal (up to a constant factor),
which is a frequent source of confusion.

Consider finally a periodic perturbation
that is switched on adiabatically, $A(t)=\exp(-i\omega t+\eta t)$,
implying again the limit $\eta\to0^+$.
The response can be written $B(t)=\chi(\omega)A(t)$,
introducing a \D{dynamic susceptibility}
\begin{equation}\label{EAchi}
  \chi(\omega) := \int_0^\infty\!{\rm d}t\,{\rm e}^{i(\omega+i\eta)t}\,R(t).
\end{equation}
This motivates the definition
\begin{equation}\label{Eft}
   F(\omega) :=
     \int_0^\infty\!{\rm d}t\, {\rm e}^{i\omega t}\,\Phi(t).
\end{equation}
of the one-sided Fourier transform $F(\omega)$
of the relaxation function $\Phi(t)$.
Because of (\ref{ERPhi}),
there is a simple relation between $\chi$ and $F$:
\begin{equation}\label{EAchiF}
   \chi(\omega) = \Phi(0) + i\omega F(\omega).
\end{equation}
In consequence, the \D{imaginary} part of the susceptibility,
which typically describes the loss peak
in a spectroscopic experiment,
is given by the \D{real} part
of the Fourier transform of the relaxation function,
$\mbox{Im~}\chi=\omega \mbox{Re~}F(\omega)$.
Conversely, \D{dispersion}\index{dispersion}
is described by 
$\mbox{Re~}\chi=\Phi(0)-\omega \mbox{Im~}F(\omega)$.

Up to this point,
the only physical input has been Eq.~(\ref{EALR}).
To make a connection with
\D{correlation functions}\index{correlation function},
more substantial input is needed.
Using the full apparatus of statistical mechanics
(Poisson brackets, Liouville equation, Boltzmann distribution, Yvon's theorem),
it is found \cite{Kub66} that for classical systems
\begin{equation}\label{EFDT}
  \langle A(t) B(0) \rangle = k_{\rm B}T \Phi(t).
\end{equation}
This is an expression of the
\D{fluctuation-dissipation theorem}\index{fluctuation-dissipation theorem}
(Nyquist 1928, Callen, Welton 1951):
the left side describes fluctuations \D{in} equilibrium;
the right side relaxation \D{towards} equilibrium,
which is inevitably accompanied by \D{dissipation}\index{dissipation}
(loss peak in $\mbox{Im~}\chi$).

Pair correlation functions are typically measured in
\D{scattering} experiments.
For instance, inelastic neutron scattering 
at wavenumber~$q$ measures the scattering law $S(q,\omega)$,
which is the Fourier transform of the density correlation function,
\begin{equation}\label{ESqw}
  S(q,\omega)=\frac{1}{2\pi}\int_{-\infty}^{\infty}\!{\rm d}t\,{\rm e}^{i\omega t}
    \langle \rho(q,t)^* \rho(q,0) \rangle.
\end{equation}
In contrast to (\ref{Eft}) and~(\ref{EAchi}),
this is a normal, two-sided Fourier transform.
In consequence,
if we let $\langle\rho(q,t)^* \rho(q,0)\rangle=\Phi(t)$,
then the scattering law $S(q,\omega)$
is proportional to the real part $\mbox{Re~}F(\omega)$
of the one-sided Fourier transform of~$\Phi(t)$.

\section{Debye's theory of dielectric relaxation}\label{ADebye}

\index{Debye relaxation}

In modern terms, 
Debye's theory of dipolar relaxation is based on a Smoluchowski equation
that describes the time evolution of
the probability distribution~$f(\vartheta,\phi,t)$
of dipole orientations $\vartheta,\phi$
($0\le\vartheta\le\pi$, $0\le\phi<2\pi$):
\begin{equation}
  \zeta \partial_t f = \nabla(\beta^{-1}\nabla f+U \nabla f)
\end{equation}
where $\zeta$ is a friction coefficient, $U$ is an external potential,
and $\beta\equiv1/(k_{\rm B}T)$.
Inserting spherical coordinates, ignoring $\phi$, and keeping $r$ constant,
we obtain
\begin{equation}\label{Esmotheta}
  \zeta' \partial_t f = (\sin\vartheta)^{-1}\partial_\vartheta \sin\vartheta
   (\beta^{-1}\partial_\vartheta f+f\partial_\vartheta U)
\end{equation}
with $\zeta'=r^2\zeta$.
An electric field in $z$ direction causes a potential
\begin{equation}
  U(t)=-\mu E(t) \cos\vartheta
\end{equation}
that is proportional to the dipole moment $\mu$.

With the ansatz
\begin{equation}
  f(\vartheta,t) = 1 + \beta g(t) \cos\vartheta,
\end{equation}
and introducing the \D{relaxation time} $\tau:=\beta\zeta'/2$,
Eq.~(\ref{Esmotheta}) simplifies to
\begin{equation}
  \tau\partial_t g(t) = -g(t)+\mu E(t) + {\cal O}(g\beta\mu E).
\end{equation}
Under realistic experimental conditions, we always have $\mu E\ll \beta^{-1}$
so that the last term is negligible.
The remaining linear differential equation shall be
rewritten for a macroscopic observable, the polarization
\begin{equation}
  P(t) = \int\!\frac{{\rm d}\Omega}{4\pi}\,\mu\cos\vartheta f(\vartheta,t)
     = \frac{\mu \beta g(t)}{3}.
\end{equation}
We obtain
\begin{equation}
  (1+\tau\partial_t) P(t) = \frac{\mu^2\beta}{3} E(t).
\end{equation}

In the simplest time-dependent experiment,
the electric field is adiabatically switched on, then suddenly switched
off at $t=0$. The polarization then
relaxes exponentially,\index{exponential relaxation}
$P(t)\propto \exp(-t/\tau)$.
In a frequency-dependent experiment, a periodic perturbation
$E(t)\propto\exp(i\omega t)$ is applied.
This yields the \D{susceptibility}\index{susceptibility}
\begin{equation}\label{EsusDeb}
  \chi_{\rm dipolar}(\omega) = \frac{P(\omega)}{E(\omega)}
     = \frac{1}{1-i\omega \tau}.
\end{equation}
The relative electric permittivity is then
\begin{equation}
   \epsilon(\omega) = 1 + \chi_{\rm dipolar}(\omega) + \chi_{\rm other}(\omega)
\end{equation}
where the ``other'' contribution comes mainly from the
electronic polarizability.
The dashed lines in Fig.~\ref{Fa-relax-havneg} show the
\D{dispersion}\index{dispersion} step in the real part $\epsilon'(\omega)$
and the \D{dissipation}\index{dissipation}
maximum in the imaginary part $\epsilon''(\omega)$.

\section{Zimm's theory of chain dynamics in solution}\label{AZimm}

\index{Zimm theory}
Starting with equation (\ref{Eentroforce}),
the entire Rouse model is based on the assumption
that the chain conformation is driven by entropy only.
This a good approximation for melts, but generally not for
solutions\index{polymer solution},
except in the $\Theta$ condition.
To account for the \D{swelling}\index{swelling} of a polymer in solution,
one needs at least to model the mutual sterical exclusion
of different chain segments.
The simplest approximation for this
\D{excluded volume interaction}\index{excluded volume}
is the repulsive potential
\begin{equation}
  U_{\rm ex}\{\v{r}\} = k_{\rm B}T v_{\rm ex} \sum_{n\ne m}\delta(\v{r}_n-\v{r}_m).
\end{equation}
As described in Sect.~\ref{SKuhn},
its effect upon the equilibrium structure
is limited to the modification of the exponent~$\nu$ in
the scaling laws (\ref{ERe}), (\ref{ERg}) for the coil radius.

For the \D{dynamics},
another modification of the Rouse model is even more important:
one has to include the
\D{hydrodynamic interaction}\index{hydrodynamic interaction}
between the polymer and the solvent.
The motion of a polymer bead drags the surrounding solvent with it,
thereby creating a flow pattern, which in turn exerts a force upon other beads.
If inertia is neglected, 
the friction term assumed in the Rouse model implies
$\v{v}=\zeta^{-1}\v{F}$.
To account for hydrodynamic interactions,
this equation must be replaced by
\begin{equation}
  \v{v}_n = \sum_m \v{H}_{nm}\v{F}_m.
\end{equation}
To estimate the coupling coefficients $\v{H}_{nm}$,
one usually refers to a simple case for which the hydrodynamic interaction
can be obtained from first principles:
for a point particle, located at $\v{r}_1(t)$
and dragged by a force $\v{F}_1$,
one can solve the Navier-Stokes equations to obtain the flow field
$\v{v}(r)=\v{H}(\v{r}-\v{r}_1)\v{F}_1$ with the
\D{Oseen tensor}\index{Oseen tensor}\footnote
{I am unable to trace back this result to C.~W.~Oseen
whose 1910 papers in Ark.\ Mat.\ Astr.\ Fys.\ are frankly unreadable.
Zimm (1956) takes the tensor from Kirkwood and Riseman (1948)
who cite a report by Burgers (1938)
that is not easily available.}
\begin{equation}
  \v{H}(\v{r}) = \frac{1}{8\pi\eta r}\left( \v{1} + \v{\hat r}\otimes\v{\hat r}
  \right),
\end{equation}
which is then used to approximate
\begin{equation}\label{EHnm}
  \v{H}_{nm} \simeq \left\{ 
  \begin{array}{ll}
    \zeta^{-1}\v{1} &\mbox{ for }n=m,\\
    \v{H}(\v{r}_n-\v{r}_m) &\mbox{ else.} 
  \end{array}\right.
\end{equation}
Unfortunately, the $\v{r}$ dependence of $\v{H}$ makes
the modified Langevin equation nonlinear.
This obstacle is overcome in the \D{Zimm theory} (1956)
by a \D{preaveraging} step
that is basically a mean field approximation:
$\v{H}_{nm}$ is replaced by its average under the equilibrium distribution
$P\{\v{r}\}$.
In the $\Theta$ condition, one obtains for $n\ne m$
\begin{equation}\label{EHnmTheta}
  \langle \v{H}_{nm} \rangle  \simeq \frac{\v{1}}{{(6\pi^3|n-m|)}^{1/2}\eta b},
\end{equation}
which is a rather long-ranged interaction.
In other conditions, a modified distribution $P\{\v{r}\}$ might be used,
leading to a modified power law $|n-m|^{-\nu}$.

The preaveraging linearization allows to rewrite 
the one-dimensional Rouse mode Langevin equation (\ref{Erouselangevin8}) with
hydrodynamic interaction as
\begin{equation}\label{Ezimmlangevin}
  \partial_t \V{\tilde{x}}
  = \M{\tilde H}
    \left( -m\partial_t^2 \V{\tilde{x}}
           - \kappa\,\M{\Lambda}\, \V{\tilde{x}} + \V{\tilde{f}} \right)
\end{equation}
with the diagonal matrix $\Lambda_{pq}=\delta_{pq}\lambda_p$
and with $\M{\tilde H}=\M{A}^{\rm T}\,\M{H}\,\M{A}$,
which
in the $\Theta$ condition is in good approximation
\begin{equation}
  \tilde{H}_{pq} \simeq \left\{ 
  \begin{array}{ll}
    0 &\mbox{ for }p\ne q,\\[1.1ex]
    \DS \frac{8 N^{1/2}}{{3(6\pi^3)}^{1/2}\eta b} &\mbox{ for }p=q=0,\\[3.4ex]
    \DS \frac{N^{1/2}}{{(3\pi^3p)}^{1/2}\eta b} &\mbox{ else.} 
  \end{array}\right.
\end{equation}

For the $p=0$ eigenmode, we find again Brownian motion,
with the Zimm diffusion constant
\begin{equation}\label{EDZimm}
  D_{\rm Z}=\frac{k_{\rm B}T H_{00}}{N} \propto N^{-1/2},
\end{equation}
which differs from $D_{\rm R}\sim N^{-1}$ in the Rouse model.
Allowing for excluded volume interaction in a good solvent,
the aforementioned generalisation of (\ref{EHnmTheta})
leads to $D_{\rm Z}\sim N^{-\nu}$.
Comparing with (\ref{ERg}), we find that the diffusion constant
is in both cases determined by the coil radius: $D_{\rm Z}\sim R^{-1}$.
This result is routinely used in
\D{photon correlation spectroscopy}\index{photon correlation spectroscopy}
(somewhat misleadingly also called 
\D{dynamic light scattering}\index{dynamic
light scattering|see{photon correlation spectroscopy}}),
where the diffusion coefficient of dilute macromolecules
is measured in order to determine their gyration radius.

For nonzero eigenmodes,
the Rouse time\index{Rouse time} is replaced in the $\Theta$ condition by 
\begin{equation}\label{EtauH}
  \tau_{\rm Z} :=\frac{\eta b^3 N^{3/2}}{(3\pi)^{1/2} k_{\rm B}T},
\end{equation}
and the mode relaxation times become
\begin{equation}
  \tau_p \simeq \frac{\tau_{\rm Z}}{p^{3/2}}.
\end{equation}
Again, the $N$ dependence can be generalized towards a dependence on
the coil radius, $\tau_{\rm Z}\sim R^3$.

\newpage

\def\ouml{\"o}
\def\uuml{\"u}

\printindex

\end{document}